\documentclass[longauth]{aa}  
\usepackage{lipsum} 
\usepackage{graphicx}
\usepackage{xcolor}
\usepackage{soul}
\usepackage{indentfirst}
\usepackage{booktabs,multirow}
\usepackage{lscape}
\usepackage{supertabular}
\usepackage{txfonts}
\usepackage{longtable}
\usepackage[colorlinks = true,
            linkcolor = blue,
            urlcolor  = blue,
            citecolor = blue,
            anchorcolor = blue]{hyperref}
\usepackage{amsmath}

\newcommand{\fesc}{$f_\mathrm{esc}^\mathrm{LyC}$}
\newcommand{\alphanth}{\ifmmode \alpha_\mathrm{nth} \else $\alpha_\mathrm{nth}$\fi}
\newcommand{\alphacs}{$\alpha^{\mathrm{3GHz}}_\mathrm{6GHz}$}
\newcommand{\Oratio}{O$_{32}$}
\newcommand{\sfrdensity}{$\Sigma_\mathrm{SFR}$}

\newcommand{\betatwo}{$\beta_{1550}$}
\newcommand{\hi}{H{\sc i}}
\newcommand{\hii}{H{\sc ii}}
\newcommand{\RCth}{RC$_\mathrm{th}$}
\newcommand{\RCnth}{RC$_\mathrm{nth}$}

\newcommand{\ewhbeta}{EW(H$\beta$)}
\newcommand{\fth}{\ifmmode f_\mathrm{th} \else f$_\mathrm{th}$\fi}
\newcommand{\nuturn}{$\nu_\mathrm{t}$}
\newcommand{\odepth}{$\tau_\nu$}
\newcommand{\SFRUV}{SFR$_\mathrm{UV}$}

\newcommand{\SFRHbeta}{SFR$_\mathrm{H\beta}$}

\newcommand{\SFRRC}[1]{ SFR$_{\mathrm{#1}}^{\mathrm{RC}}$ }
\newcommand{\mstar}{$\mathrm{M_*}/\mathrm{M_\odot}$}

\newcommand{\revtext}[1]{#1}

\begin{document}

   \title{Low-redshift Lyman Continuum Survey (LzLCS):}

   \subtitle{Radio continuum properties of low-$z$ Lyman continuum emitters}

   \author{
          Omkar Bait 
          \inst{1}
          \fnmsep\thanks{O. Bait is the corresponding author}
          \and
          Sanchayeeta Borthakur
          \inst{2}
          \and
          Daniel Schaerer
          \inst{1,3}
          \and
          Emmanuel Momjian
          \inst{4}
          \and
          Biny Sebastian 
          \inst{5}
          \and
          Alberto Saldana-Lopez
          \inst{25}
          \and 
          Sophia R. Flury
          \inst{6}
          \and
          John Chisholm
          \inst{7}
          \and
          Rui Marques-Chaves
          \inst{1}
          \and
          Anne E. Jaskot
          \inst{8}
          \and
          Harry C. Ferguson
          \inst{9}
          \and
          Gabor Worseck
          \inst{10}
          \and
          Zhiyuan Ji
          \inst{11}
          \and
          Lena Komarova
          \inst{12}
          \and
          Maxime Trebitsch
          \inst{13}
          \and
          Matthew J. Hayes
          \inst{14}
          \and
          Laura Pentericci
          \inst{15}
          \and
          Goran Ostlin
          \inst{14}
          \and
          Trinh Thuan
          \inst{16}
          \and
          Ricardo O. Amor\'{i}n
          \inst{17, 18}
          \and
          Bingjie Wang
          \inst{19, 20, 21}
          \and
          Xinfeng Xu
          \inst{22, 23}
          \and
          Mark T. Sargent
          \inst{24}
          }

   \institute{Observatoire de Gen\`eve, Universit\'e de Gen\`eve, Chemin Pegasi, 1290 Versoix, Switzerland\\
              \email{omkar.bait@unige.ch}
         \and
             School of Earth and Space Exploration, Arizona State University, 781 Terrace Mall, Tempe, AZ 85287, USA
        \and CNRS, IRAP, 14 Avenue E. Belin, 31400 Toulouse, France
        \and
             National Radio Astronomy Observatory, P.O. Box O, Socorro, NM 87801, USA
        \and
             Department of Physics and Astronomy, University of Manitoba, Winnipeg, Canada
        \and
             Department of Astronomy, University of Massachusetts Amherst, Amherst, MA 01002, United States 
        \and
             Department of Astronomy, The University of Texas at Austin, 2515 Speedway, Stop C1400, Austin, TX 78712-1205, USA
        \and
             Department of Astronomy, Williams College, Williamstown, MA 01267, United States
        \and
             Space Telescope Science Institute, 3700 San Martin Drive, Baltimore, MD, 21218, USA
             \and
             VDI/VDE Innovation + Technik GmbH
             \and
             Steward Observatory, University of Arizona, 933 N. Cherry Avenue, Tucson, AZ 85721, USA
             \and
             Department of Astronomy, University of Michigan, 1085 S. University, Ann Arbor, MI 48109, USA
             \and
             Kapteyn Astronomical Institute, University of Groningen, P.O. Box 800, 9700 AV Groningen, The Netherlands
             \and
             Stockholm University, Department of Astronomy and Oskar Klein Centre for Cosmoparticle Physics, AlbaNova University Centre, SE-10691, Stockholm, Sweden
             \and
             INAF- Osservatorio, Astronomico di Roma, Via di FRascati, 33 00078, Monte Porzio Catone (Italy)
             \and
             Astronomy department, University of Virginia, P.O. Box 400325, Charlottesville, VA 22904-4325, USA 
             \and
             ARAID Foundation. Centro de Estudios de F\'{\i}sica del Cosmos de Arag\'{o}n (CEFCA), Unidad Asociada al CSIC, Plaza San Juan 1, E--44001 Teruel, Spain
             \and
             Departamento de Astronom\'{i}a, Universidad de La Serena, Av. Juan Cisternas 1200 Norte, La Serena 1720236, Chile
             \and
             Department of Astronomy \& Astrophysics, The Pennsylvania State University, University Park, PA 16802, USA
             \and
             Institute for Computational \& Data Sciences, The Pennsylvania State University, University Park, PA 16802, USA
             \and
             Institute for Gravitation and the Cosmos, The Pennsylvania State University, University Park, PA 16802, USA
             \and
             Department of Physics and Astronomy, Northwestern University, 2145 Sheridan Road, Evanston, IL, 60208, USA.
             \and 
             Center for Interdisciplinary Exploration and Research in
             Astrophysics (CIERA), Northwestern University, 1800 Sherman Avenue, Evanston, IL, 60201, USA.
             \and
             International Space Science Institute (ISSI), Hallerstrasse 6, CH-3012 Bern, Switzerland
             \and 
             Department of Astronomy, Stockholm University, Oscar Klein Centre, AlbaNova University Centre, 106 91 Stockholm, Sweden
             }

\date{Received ; accepted }

 
  \abstract
 { Sources that leak Lyman continuum (LyC) photons and lead to the reionisation of the universe are an object of intense study using multiple observing facilities. Recently, \revtext{the} Low-redshift LyC Survey (LzLCS)  \revtext{has presented} the first large sample of LyC emitting galaxies at low redshift ($z\sim 0.3$) with the Hubble Space Telescope Cosmic Origins Spectrograph. The LzLCS sample contains a robust estimate of the LyC escape fraction (\fesc{}) for 66 galaxies, spanning a wide range of \fesc{} values.
 }
   {
Here, we aim to study the dependence of \fesc{} on the radio continuum (RC) properties of LzLCS sources.
Overall, RC emission can provide unique insights into the role of supernova feedback, cosmic rays (CRs), and magnetic fields from its non-thermal emission component. RC emission is also a dust-free tracer of the star formation rate (SFR) in galaxies.
   }
    {
    In this study, we present Karl G. Jansky Very Large Array (VLA) RC observations of the LzLCS sources at gigahertz (GHz) frequencies.  We performed VLA C (4-8 GHz) and S (2-4 GHz) band observations for a sample of 53 LzLCS sources. We also observed a sub-sample of 17 LzLCS sources in the L (1-2 GHz) band. We detected RC from both C- and S-bands in 24 sources for which we are able to estimate their radio spectral index across 3-6 GHz, denoted as \alphacs{}. We also used the RC luminosity to estimate their SFRs.
    }
   {
   The radio spectral index of LzLCS sources spans a wide range, from flat ( $\geq -0.1$) to very steep ($\leq -1.0$). They have a steeper mean \alphacs{}~(~$\approx -0.92$) compared to that expected for normal star-forming galaxies (\alphacs{}~$\approx -0.64$). They also show a larger scatter in \alphacs{} ($\sim$0.71) compared to that of normal star-forming galaxies ($\sim$0.15). 
   The strongest leakers in our sample show flat \alphacs{}, weak leakers have \alphacs{} close to normal star-forming galaxies and non-leakers are characterized by steep \alphacs{}.
   We argue that a combination of young ages, free-free absorption, and a flat cosmic-ray energy spectrum can altogether lead to a flat \alphacs{} for strong leakers. Non-leakers are characterized by steep spectra which can arise due to break or cutoff at high frequencies. Such a cutoff in the spectrum can arise in a single injection model of CRs characteristic of galaxies which have recently stopped star-formation. The dependence of \fesc{} on \alphacs{} (which is orientation-independent) suggests that the escape of LyC photons is not highly direction-dependent at least to the first order.
   The radio-based SFRs (SFR$^{\mathrm{RC}}$) of LzLCS sources show a large offset ($\sim0.59$ dex) from the standard SFR$^{\mathrm{RC}}$ calibration. We find that adding \alphacs{} as a second parameter helps us to calibrate the SFR$^{\mathrm{RC}}$ with SFR$_\mathrm{UV}$ and SFR$_\mathrm{H\beta}$ within a scatter of $\sim 0.21$ dex.
   }
   {For the first time, we have found a relation between \alphacs{} and \fesc{}. This hints at the interesting role of supernovae feedback, CRs, and magnetic fields in facilitating the escape (alternatively, and/or the lack) of LyC photons.
   }

   \keywords{galaxies: starburst -- Radio continuum: galaxies -- Radio continuum: ISM
               }
   
%

\maketitle

\section{Introduction}
\label{sec:introduction}

\revtext{Understanding how and when the Universe underwent reionisation requires knowledge of the nature of the sources that provide the ionising UV photons}. Although quasars are very bright in the UV, their number density at $z \geq 4$ is likely too low to contribute significantly to the reionising budget of the Universe \citep[see discussions in][]{Kulkarni19,Giallongo2015,Grazian2020}. However, at high redshift, low-mass star-forming galaxies dominate the luminosity (mass) functions \citep[e.g.,][]{ouchi09,Bouwens15, Grazian15, Song16} which can be significant contributors to reionising the Universe \citep{Robertson2010,Ishigaki18,  Finkelstein19}. Bright galaxies could also contribute to reionisation, as discussed by several authors in previous works \citep[see e.g.,][]{Sharma2016The-brighter-ga, Naidu20, Marques-Chaves21, Marques-Chaves2022An-extreme-blue}.

The escape of Lyman continuum (LyC) photons from galaxies before the Universe was completely reionised cannot be directly studied since the LyC photons will \revtext{be} absorbed by the intergalactic medium (IGM) \citep{Inoue14}. Thus, we have to rely on local (low redshift) analogs of high-redshift galaxies to measure the escape of LyC photons where the IGM is mostly transparent to ionizing photons. Moreover,  low redshift galaxies are easier to study in detail. Typically these local analogs ($z \sim 0.3$) are extremely young starbursts with low dust attenuation, characterized by the slope of the non-ionizing UV continumm, $\beta \leq -2$ given by $f_\lambda \propto \lambda^\beta$ \revtext{\citep[e.g.,][]{Chisholm22}}. They also display a high star-formation rate density (\sfrdensity{} $\geq 0.1$ M$_\odot$yr$^{-1}$kpc$^{-2}$) and high ionization ratio ([OIII]/[OII]; \Oratio $\geq 3$), along with  several other extreme emission line properties  \citep[e.g.,][]{cardamone2009, yang2017, Izotov11, Izotov21a}. These galaxies can efficiently produce ionising photons \citep{Schaerer16} and also leak a large amount of them \citep{Izotov16a, Izotov16b, Izotov18a,Izotov18b, Izotov21b}; thus, their counterparts at high-$z$ can significantly contribute to reionisation. Recent James Webb Space Telescope (JWST) observations have confirmed that these are excellent local analogs of $z > 6$ galaxies (i.e., during the epoch of reionisation) since they show several similar extreme emission line properties \citep[e.g.,][]{Schaerer22,Rhoads23, Matthee23, Cameron23, Mascia23}.

A systematic study of low redshift LyC emitters (LCEs, often also referred to as LyC leakers) was lacking until recently. The high success rate in finding LCEs in low-$z$ analogs has motivated the Low-$z$ Lyman Continuum Survey \citep[][LzLCS, PI: Jaskot HST Project ID: 15626]{Flury22a}. This has enabled us to measure the escape fraction of LyC photons (\fesc{}) in a statistically large sample of galaxies. Here, \fesc{} is defined as the ratio between the observed and the intrinsic ionizing fluxes and it is crucial parameter to constrain various models of reionization. LzLCS is a systematic survey of 66 star-forming galaxies at $z \sim 0.3$ using \revtext{Hubble Space Telescope (HST) Cosmic Origins Spectrograph (COS)} to measure their LyC. They are selected 
from the Sloan Digital Sky Survey (SDSS) and Galaxy Evolution Explorer (GALEX), with a range of \Oratio{}, star-formation rate densities (\sfrdensity) and UV-slopes $\beta$.
The LzLCS has found strong ($ > 2\sigma$) LyC detection in 35 galaxies with a range of LyC escape fractions (\fesc{} up to $\sim$ 50\%), and has been combined with previous LyC measurements to provide a sample of 89  low-$z$ galaxies with LyC measurements.
Such a sample can then be used to study the dependence of various UV/optical-based physical properties on \fesc{} \citep{Flury22b, Saldana-Lopez22, Marques-Chaves22b, Chisholm22, Wang21, Xu23}.   

The exact physical mechanisms which lead to the leakage of LyC photons are still a matter of debate. LyC photons can either escape through low \hi{} column density holes carved in the interstellar medium (ISM) \citep{Gazagnes18, Saldana-Lopez22} or from a density-bounded nebula \citep[e.g.,][]{Nakajima14, Zackrisson13, Bremer23}. Simulations show that strong supernova (SN) driven feedback can play an important role in the escape of LyC photons \citep{Kimm14, Paardekooper15, Kimm17, Trebitsch17, Ma20, Barrow20}. \citet{Kakiichi21} also {show} the effect of turbulent ISM on the escape of LyC photons. Recently \citet{Katz23} showed that escape of LyC photons can occur from different modes of star formation. \citet{Ma20} used  zoom-in cosmological simulations to find that most LCEs are associated with a feedback-driven kiloparsec-scale superbubble and host very young stellar populations ($\la 10$ Myr). Observationally, this can be seen, for example, in the Sunburst arc, where the regions that are leaking exhibit a very young stellar population and outflow signatures, as opposed to the non-leaking regions \citep{Mainali22, Kim23}. \revtext{Simulations have also shown rapid fluctuations in \fesc{} (on 10 Myr timescales) due to the stochastic and bursty nature of star formation} \citep[e.g.,][]{Ma15,Trebitsch17, Kimm17, Barrow20, Endsley23}. All these effects complicate the physical understanding and predictability of \fesc{} using various diagnostics.

Radio observations provide a complimentary view into these galaxies.  For example, the radio continuum (RC) is a dust-free tracer of the total star-formation rate (SFR) in galaxies \citep{yun2002, murphy2011}. For normal galaxies, the RC at 1.4 GHz and other mid-gigahertz (mid-GHz) frequencies (3, 6, and 10 GHz) has been extensively used to measure the SFR of galaxies both nearby and at high redshifts \citep[e.g.,][]{Kennicutt12, tabatabaei2017}. Modelling the radio-spectral energy distribution (radio-SED), especially across several widely separated frequencies, enables us to study several unique physical parameters \citep[][]{tabatabaei2017}; for example, the supernova rate, the role of cosmic rays (CRs), emission measure based on the free-free absorption optical depth \citep{condon1992, Hunt04}, and the equipartition magnetic field \citep{beck2005, beck2015}.

The RC emission at GHz frequencies in normal galaxies arises due to a combination of non-thermal synchrotron emission, which approximately follows a power-law with a slope of \alphanth{} $\sim -0.8$ and a thermal free-free component with a power law slope of $-0.1$ \citep{Condon91}.
In more complex scenarios the radio-SED shows a deviation from a simple power law behaviour. For example, in the case of dense starbursts in ultraluminous infrared galaxies (ULIRGS), blue-compact dwarfs (BCDs), and metal-poor dwarf galaxies, it has been observed that the radio-SED shows a turnover below $\sim~1-2$ GHz \citep[e.g.,][]{Condon91, 
Hunt04}.
Such a turnover in dense starbursts can arise from free-free absorption (FFA) due to large optical depths. Moreover, the RC spectrum can also show multiple FFA components \citep[e.g.,][]{Clemens10, Galvin18}. In addition at mid-to-high frequencies, they can show a much steeper spectrum than the canonical slope of -0.7, due to a steeper non-thermal spectral index \citep{Galvin18}. Deviations in the non-thermal spectral index (\alphanth{}) are also observed in local BCDs, showing both steeper and flatter \alphanth{} than normal galaxies, possibly due to the inhomogeneous nature of star-formation in these galaxies \citep[e.g.,][]{Klein84, Klein91, Deeg93}. Steep \alphanth{} values are associated with a general lack of CR confinement \citep{Skillma88, Klein91}. The exact physical mechanisms for this steepening are not well known, but it can be due to synchrotron ageing particularly in a strong magnetic field, inverse-Compton (IC) losses or even the escape of CRs \citep[see][ for a discussion]{Klein91, ramya2011}. Such deviations have systematic effects on the radio-SED and can be used to distinguish different physical processes \citep[see e.g.,][]{klein2018, Clemens10, Galvin18}. Overall the radio spectral index can also be used as a proxy for the age of starburst \citep{Cannon04, Hirashita06}. In young, metal-poor, and dense starbursts \citep[e.g., SBS 0335-052;][]{Hunt04}, the non-thermal emission is associated with an ensemble of compact SN remnants (SNRs) that are expanding in a dense  ISM \citep[e.g.,][]{Hunt04, Hirashita06} instead of diffuse CRs in normal galaxies. 

Of particular interest is the non-thermal luminosity which (along with the non-thermal spectral index) can be used to estimate the Type II SN rate \citep{condon1992}. This can help us in understanding the role of SN feedback and, thus, also indirectly the contribution of SN to the escape of ionizing photons. 
The radio spectral index and the deviations between radio and other UV/optical/IR tracers are also used to understand changes in the recent star-formation history \citep[e.g.,][]{Cannon04, bressan2002, Hirashita06, Arango-Toro23}. There is no systematic RC study of LCEs to date. \revtext{Past} RC studies of local LCEs on Haro 11 \citep{Hayes07} and Tol1247-232 \citep{Puschnig17} found a lower radio-based SFR compared to the SFR from other tracers. Furthermore, RC studies of sources most closely resembling LCEs, for instance green peas (GPs) and blueberries (BBs) also have suppressed radio-based SFR compared to other SFR indicators \citep{chakraborti2012, Sebastian19}. It is speculated that such a suppression could be because galaxies with young stellar populations do not have any SN activity yet; thus, \revtext{they} would  have a lower non-thermal emission component than normally expected \citep{Hayes07, chakraborti2012, Sebastian19}. Moreover, the loss of CRs either via diffusion or outflows might also be playing a role in causing the deficit \citep{Sebastian19}. Such a deficit has also been observed in local analogs of Lyman-break galaxies (LBGs) and for $z \sim 3$ LBGs based on a stacking analysis \citep{carilli2008, greis2017}.   The exact causes for this deficit are not yet completely investigated due to their unknown radio-SED. Several of the factors mentioned above for powerful starbursts are also valid for GPs and BBs which can lead to these variations.

In this study, for the very first time, we present RC observations of low-$z$ star-forming galaxies with LyC measurements, which allows us to examine if and how their RC emission relates to the escape of ionizing photons. For instance, the radio thermal fraction and spectral index depend on the past SN activity and star-formation history (SFH) of galaxies. Importantly, the radio spectral index can be used as a proxy for the age of the starburst \citep{Cannon04, Hirashita06}. The dependence of these radio properties on \fesc{} can tell us, for example, about the role of SN-driven feedback and SFH of galaxies on the escape of LyC photons. To do so, we target $z \sim 0.3-0.4$ galaxies from the LzLCS sample. Our observations were conducted using the NRAO\footnote{The National Radio Astronomy Observatory is a facility of the National Science Foundation operated under cooperative agreement by Associated Universities, Inc.} Karl G. Jansky Very Large Array (VLA) for a sample of 53 LzLCS sources at 6 (C-band), 3 (S-band), and  (a subsample) at 1.5 (L-band) GHz. 

Our paper is organized as follows. In Section \ref{sec: data} we describe our observation setup, data analysis and detection, and non-detection statistics. In Section \ref{sec: results}, we present the radio spectral index (across 3-6 GHz; \alphacs{})  and thermal fraction distribution, and radio-SED for our sample of detections. Our main result is described in Section \ref{sec: specindex vs fesc} where we show a dependence between \alphacs and \fesc{}. In Section \ref{sec: discussion}, we discuss the various physical reasons that can give rise to such a relation. In Section \ref{sec: conclusions}, we summarize our main results and conclude. Throughout this paper, we assume $H_0=70$ km s$^{-1}$\ Mpc$^{-1}$, $\Omega_m=0.3$, and $\Omega_\Lambda=0.7$, which is same as that used in \citet{Flury22a}.

\section{Data}
\label{sec: data}
In this section, we describe our sample selection and VLA observations at multiple observing bands, namely: the C (4-8 GHz), S (2-4 GHz), and L (1-2 GHz) with the B-configuration (PI: Borthakur; Project Code: 21B-111). We also describe our calibration and imaging strategy. 

\subsection{Sample selection}
Our current sample consists of 53 LzLCS sources observed with the VLA. This sample is a subset of the complete 66 galaxies from LzLCS. 
The LzLCS sample spans a wide range in \fesc{} and can hence be used to do a statistical study of LCEs and its dependence on other physical properties \citep{Flury22b, Saldana-Lopez22, Marques-Chaves22b, Xu23}. Our 53 LzLCS sources with VLA observations were selected on the basis of their expected flux densities at L-band being above 40 $\mu$Jy at 5-$\sigma$. This corresponds to a SFR of 6.6 $\mathrm{M}_\odot$/yr for the median redshift ($z \sim 0.3$) of our sample based on the standard radio-SFR relation \citep{murphy2011}. 
The various physical properties of the LzLCS sources studied here are taken from \citet{Flury22a} and \citet{Saldana-Lopez22}.

\subsection{Observation setup}
\label{sec:obs_setup}

We used the default VLA setup for our C-, S-,  and L-  band continuum observations. For the C-band (4-8 GHz), we used a three-bit sampler with a total bandwdith of 4 GHz. For the S-band (2-4 GHz) and L-band (1-2 GHz), we used the eight-bit sampler with a total bandwidth of 2 and 1 GHz respectively. For the C-band setup, we split the entire bandwidth into two basebands each having 16 $\times$ 128 MHz subbands. Each subband was further split into 64 channels. Similarly for the S-band setup, we split the entire bandwidth into two basebands each having 8 $\times$ 128 MHz subbands where each subband was further split into 64 channels. And for the L-band setup, we split the entire bandwidth into two basebands each having 8 $\times$ 64 MHz sub-bands where each sub-band was further split into 64 channels. 
All our observations were conducted in the B-configuration with a few in the BtoBnA, BnA, and BnAtoA move time, depending on the scheduling constraints. We were thus able to typically achieve an angular resolution of 1.6\arcsec{}, 3.2\arcsec{} and 6.9\arcsec{} for C-, S-, and L-bands, respectively.


We followed the standard observing strategy used for continuum observations. We typically spent a minute on the initial setup and then 3-5 minutes on the flux density scale calibrator. We then observed the phase calibrator for a minute and the target for 5-6 minutes. We then cycled between the phase calibrator and the target. We had a total on-source time of 22 minutes for the C- and S-band observations, as well as a total on-source time of 75 minutes for the L-band observations. The C- and S-band observations of a given target source were combined in a single $\sim$ 1\,hr long scheduling block (SB), while their L-band observations were in separate SB with a length of $\sim$1\,hr and 40\,min.
Table \ref{table: obs detail} and \ref{table: obs detail L band} summarizes the S-, C- and L-band observing details of our sample. In total, we were able to target 53 sources in both the S- and C-bands, and a subset of 17 in the L-band.

\subsection{Data analysis: Calibration and imaging }
\label{sec:data_analysis}
We used the Common Astronomy Software Applications (\textsc{CASA}) data processing software \citep{mcmullin2007casa, CASA} for all our data analysis. Each SB containing the multi-band C- and S-band data was initially flagged and calibrated using the VLA calibration pipeline v6.2.1. The calibrated flux density scale and phase calibrators for each SB were manually inspected for any leftover RFI and poorly performing antennas. These were then flagged and the datasets were re-run using the latest CASA pipeline v6.4.1. For some datasets, several rounds of the pipeline were run, after having performed flagging in each round. 

The final calibrated target data were then extracted and any leftover RFI in some spectral windows was flagged using the task \textsc{rflag}. We then imaged these data using CASA task \textsc{tclean}, separately by splitting the target in C- and S-bands. We use the multi-term multi-frequency synthesis (mtmfs) algorithm \citep{Rau11} with \textsc{nterms}=2, and a \textsc{robust=0.5}. 
We then manually made elliptical regions around each of our target sources using CASA task \textsc{imview}. We then estimated the flux densities of these sources using 2D Gaussian fit using the task \textsc{imfit}. We used this same region to estimate the noise in the images from a emission free area. For the one resolved source in our sample (J095700+235709), we estimated the fluxes in different bands and the corresponding errors using PyBDSF \citep{Mohan15}. The estimated flux densities and noise ($\sigma$) for our sources in C-, S-, and L-bands are presented in Table \ref{table: radio properties}. For sources with a signal-to-noise ratio (S/N) below 3, we quote the 3$\sigma$ upper limits in the flux density. In Table \ref{table: detections stats} we present a summary of the number of detections and non-detections in different bands. In all further analysis we add a 5\% calibration error \citep[e.g.,][]{Shao22} in quadrature to the noise presented in Table \ref{table: radio properties}.

\begin{table*}[ht!]
\caption{Radio properties of our LzLCS sample.} 
\begin{tabular}{ccccc}
\hline\hline \\
\label{table: radio properties}
Object & Flux density at $3$ GHz & Flux density at 6 GHz & Flux density at 1.5 GHz & \alphacs{} \\  
       & ($\mu$Jy) & ($\mu$Jy) & ($\mu$Jy) &\\ \\
\hline \\
J004743+015440 & $<28.5$ & $24.0 \pm 5.5$ & $<90.0$ & $<-0.25$ \\
J011309+000223 & $<31.2$ & $--^{a}$ & $<63.0$ & $-$ \\
J012910+145935 & $<45.0$ & $75.6 \pm 7.0$ & $<63.0$ & $<0.75$ \\
J072326+414608 & $87.2 \pm 8.1$ & $<16.5$ & $<42.0$ & $>-2.4$ \\
J081112+414146 & $<44.4$ & $<19.5$ & $<42.0$ & -- \\
J081409+211459 & $142.0 \pm 8.8$ & $114.3 \pm 5.2$ & $300.0 \pm 8.0$ & $-0.313 \pm 0.033$ \\
J082652+182052 & $69.4 \pm 8.9$ & $37.0 \pm 6.6$ & $<27.0$ & $-0.907 \pm 0.209$ \\
J083440+480541 & $68.2 \pm 7.8$ & $69.1 \pm 5.1$ & $143.0 \pm 15.0$ & $0.019 \pm 0.003$ \\
J090918+392925 & $<25.5$ & $<17.4$ & $<45.0$ & -- \\
J091113+183108 & $69.2 \pm 8.5$ & $52.7 \pm 5.0$ & $<90.0^b$ & $-0.393 \pm 0.067$ \\
{J091207+523960}$^a$ & $--$ & $--$ & $<123.0$ & -- \\
J091208+505009 & $<26.1$ & $40.0 \pm 5.7$ & $<45.0$ & $<0.62$ \\
J091703+315221 & $55.6 \pm 8.2$ & $47.5 \pm 5.8$ & $80.0 \pm 20.0$ & $-0.227 \pm 0.046$ \\
J092552+395714 & $<25.8$ & $<16.5$ & $<33.0$ & -- \\
J094001+593244 & $63.0 \pm 10.4$ & $26.0 \pm 4.9$ & -- & $-1.277 \pm 0.332$ \\
J095236+405249$^b$ & $392.3 \pm 7.5$ & $50.9 \pm 5.4$ & $<36.0$ & $-2.946 \pm 0.38$ \\
{J095700+235709} & $930.0 \pm 81.2$ & $210.0 \pm 23.8$ & $1730.0 \pm 150.0$ & $-2.147 \pm 0.343$ \\
J095838+202508 & $<25.2$ & $<16.5$ & $<30.0$ & -- \\
J101401+523251 & $<22.5$ & $<16.2$ & -- & -- \\
J102615+633308 & $<25.2$ & $<15.6$ & -- & -- \\
J103344+635317 & $73.8 \pm 8.0$ & $80.7 \pm 6.0$ & -- & $0.129 \pm 0.019$ \\
J103816+452718 & $132.4 \pm 8.9$ & $40.4 \pm 5.8$ & -- & $-1.712 \pm 0.297$ \\
J105117+474357 & $<27.0$ & $<16.8$ & -- & -- \\
J105331+523753 & $158.7 \pm 8.2$ & $79.8 \pm 5.6$ & -- & $-0.992 \pm 0.111$ \\
J110452+475204 & $<23.7$ & $<16.2$ & -- & -- \\
J112224+412052 & $<25.5$ & $<17.1$ & -- & -- \\
J112848+524509 & $<23.1$ & $<16.2$ & -- & -- \\
J112933+493525 & $<22.2$ & $<18.0$ & -- & -- \\
J113304+651341 & $<24.0$ & $27.9 \pm 5.8$ & -- & $<0.22$ \\
J115855+312559 & $136.7 \pm 7.9$ & $72.9 \pm 5.6$ & -- & $-0.907 \pm 0.108$ \\
J115959+382422 & $<23.7$ & $<17.1$ & -- & -- \\
J120934+305326 & $120.7 \pm 7.8$ & $70.5 \pm 5.5$ & -- & $-0.776 \pm 0.096$ \\
J121915+453930 & $<25.5$ & $<19.2$ & -- & -- \\
J123519+063556 & $<28.5$ & $<19.5$ & -- & -- \\
J124423+021540 & $225.6 \pm 9.2$ & $145.7 \pm 5.9$ & -- & $-0.631 \pm 0.057$ \\
J124619+444902 & $61.0 \pm 12.6$ & $41.5 \pm 5.0$ & -- & $-0.556 \pm 0.139$ \\
J124835+123403 & $126.0 \pm 7.7$ & $76.0 \pm 6.0$ & -- & $-0.729 \pm 0.089$ \\
J124911+464535 & $57.0 \pm 7.9$ & $<16.8$ & -- & $>-1.76$ \\
J125718+410221$^b$ & $82.2 \pm 9.7$ & $26.6 \pm 5.5$ & -- & $-1.628 \pm 0.404$ \\
{J130128+510451 $^c$} & $488.0 \pm 8.0$ & $<16.5$ & -- & $>-4.89$ \\
J131037+214817 & $<24.3$ & $41.3 \pm 5.6$ & -- & $<0.77$ \\
{J131419+104739} & $125.0 \pm 8.3$ & $37.7 \pm 6.0$ & -- & {$-1.729 \pm 0.322$} \\
{J131904+510309} & {$98.5 \pm 8.7$} & $30.4 \pm 5.0$ & -- & {$-1.696 \pm 0.339$} \\
J132633+421824 & $95.0 \pm 8.4$ & $<17.1$ & -- & $>-2.47$ \\
J132937+573315 & $132.1 \pm 8.2$ & $29.2 \pm 5.6$ & -- & $-2.178 \pm 0.465$ \\
J134559+112848 & $249.6 \pm 7.7$ & $71.4 \pm 5.7$ & -- & $-1.806 \pm 0.2$ \\
J141013+434435$^b$ & $233.0 \pm 8.8$ & $99.6 \pm 5.4$ & -- & $-1.226 \pm 0.119$ \\
J144010+461937 & $158.9 \pm 8.3$ & $96.7 \pm 5.7$ & -- & $-0.717 \pm 0.076$ \\
J151707+370512 & $<30.0$ & $30.7 \pm 6.4$ & -- & $<0.03$ \\
J155945+403325 & $74.8 \pm 8.3$ & $87.2 \pm 5.3$ & -- & $0.221 \pm 0.032$ \\
J160437+081959 & $113.6 \pm 9.4$ & $<18.0$ & -- & $>-2.66$ \\
J164607+313054 & $<26.7$ & $<19.8$ & -- & -- \\
J164849+495751 & $<39.9$ & $<38.4$ & -- & -- \\
\hline
\end{tabular}
\tablefoot{ For non-detections in the different bands, we present the $3\sigma$ upper limits on the flux density. For estimating the upper/lower limits on \alphacs{}, we use the corresponding  $3\sigma$ upper limits on the flux density. $^{a}$ Here the noise is high due to a very strong source in the field. $^{b}$ For these sources the flux density measurement is uncertain due to artefacts from a strong source in the field. $^{c}$ For this source the flux is possibly contaminated due to AGN jet from a neighbouring source. See Appendix \ref{appendix sec: J130128_AGN_jet} for more details. }
\end{table*}

  \begin{table}[ht!]
      \caption[]{Detection and non-detection statistics for our sample of 53 LzLCS sources.}
         \label{table: detections stats}
     $$ 
         \begin{array}{p{0.8\linewidth}l}
            \hline
            \noalign{\smallskip}
            Detection/Non-detection Type      &  Number \\
            \noalign{\smallskip}
            \hline
            \noalign{\smallskip}
            Both C- \& S-band detections & 24     \\
            Both C- \& S-band non-detections & 18 \\
            C-band only non-detections & 5 \\
            S-band only non-detections & 6 \\
            \noalign{\smallskip}
            \hline
         \end{array}
     $$ 
   \end{table}

\subsection{VLA images}

\label{sec: vla maps}
\begin{figure}
  \resizebox{\hsize}{!}{\includegraphics{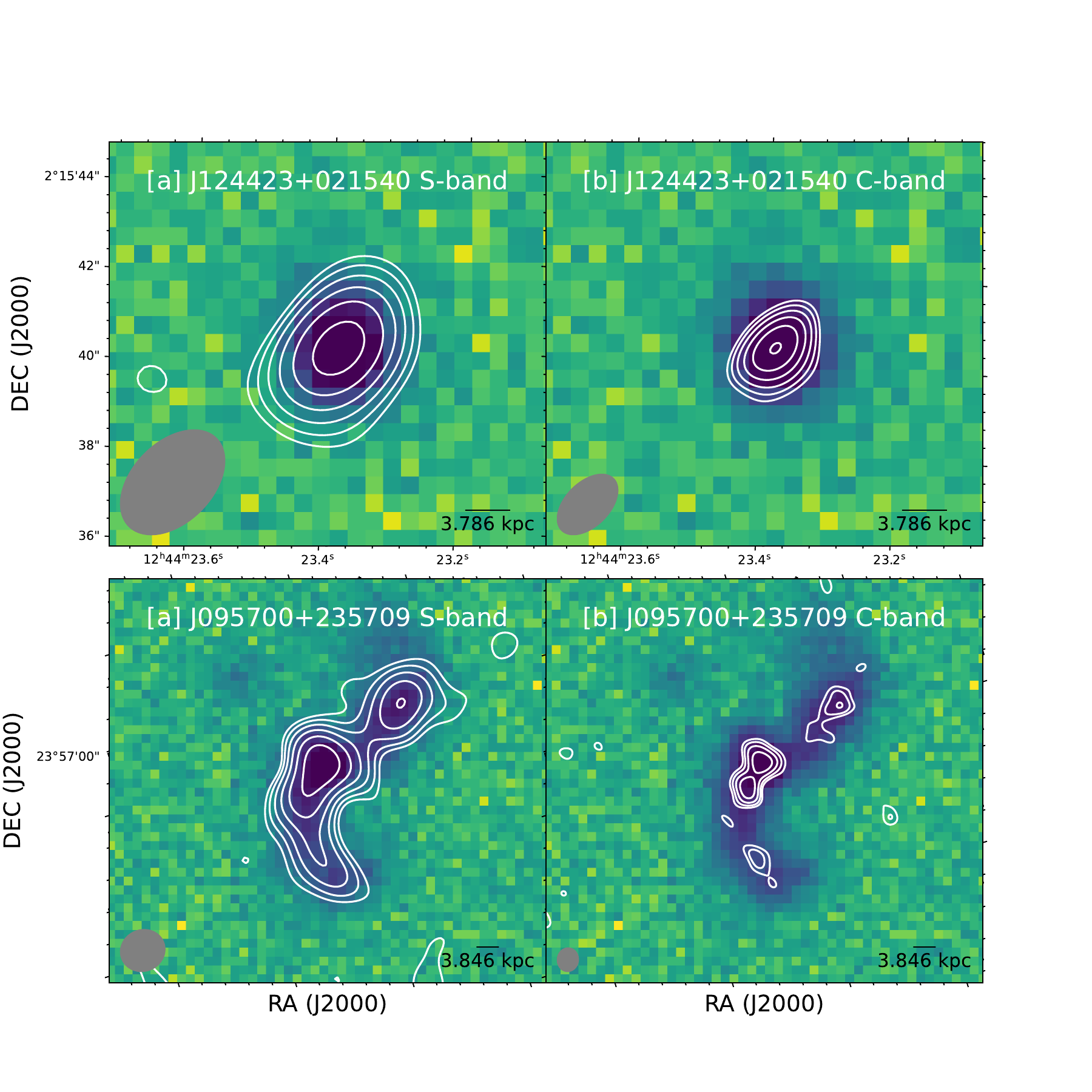}}
  \caption{ VLA images of two sources in our sample, J$124423+021540$ (top panels) and J$095700+235709$ (bottom panels), shown in white contours with the SDSS r-band image \citep{Blanton17} in the background. On the left (right) panels we show the VLA S-band (C-band) image contours. The lowest contours in each panels corresponds to 3$\sigma$ level and they rise in powers of $\sqrt{2}$.  $\sigma$ values correspond to $\sim 8$ and $\sim 5~\mu$Jy/beam for S- and C-band respectively}
  \label{fig: vla maps}
\end{figure}
In Fig. \ref{fig: vla maps}, upper panel (lower panel), we overlay the S- and C-band radio contours on SDSS r-band images in the background for J124423+021540 (J095700+235709). In Appendix \ref{appendix sec: VLA maps}, we show the C- and S-band VLA images for the rest of the detection sample. We note that except for J095700+235709, shown in Fig.~\ref{fig: vla maps}, all the sources in our sample are unresolved in both radio and optical wavelengths.





%

\section{Results}
\label{sec: results}


\subsection{Radio-SED at mid GHz frequencies}
\begin{figure}
  \resizebox{\hsize}{!}{\includegraphics{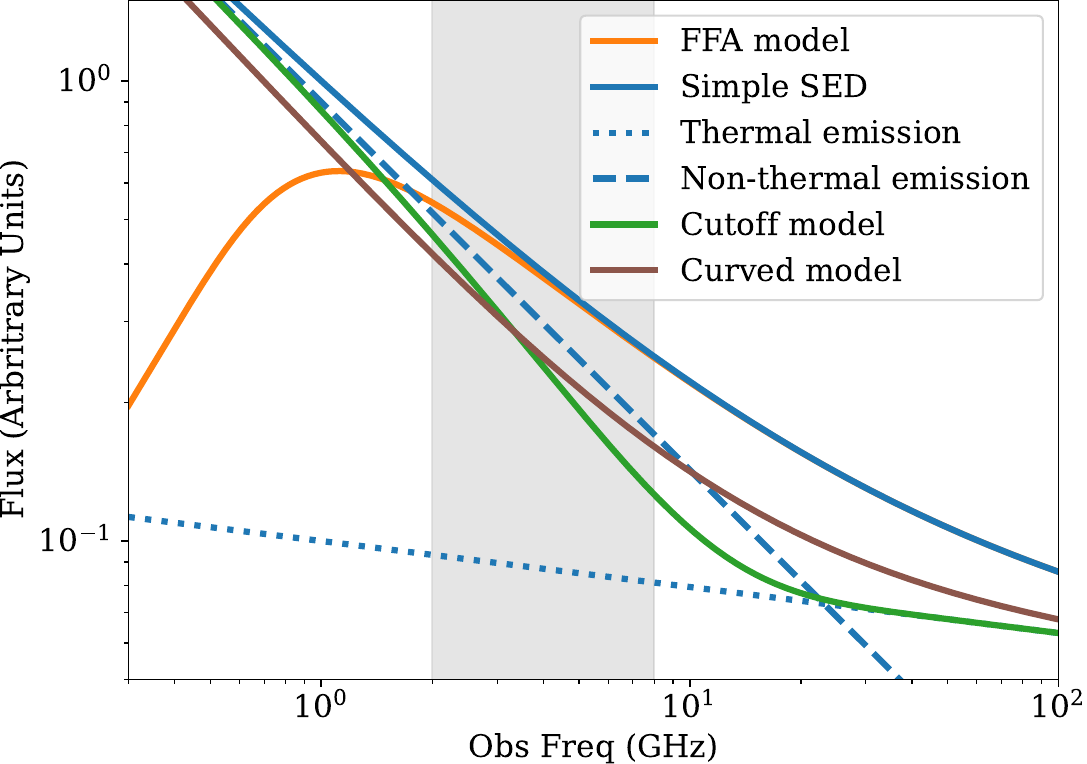}}
  \caption{{Typical radio-SED with a mixture of thermal and non-thermal emission (Eq. \ref{eq: simple sed model}) in blue line.} The corresponding thermal and non-thermal emission components are shown in blue dashed and dotted lines. We also show the effect of FFA (Eq. \ref{eq: FFA model}) with a turnover at 1 GHz in orange. And the cutoff and curved model \citep[from][]{klein2018} with a high-frequency break at 6 GHz in green and red respectively. The grey-shaded area shows the region of the SED explored in this work.
  }
  \label{fig: radio sed cartoon}
\end{figure}

\begin{figure*}
  \resizebox{\hsize}
  {!}{\includegraphics{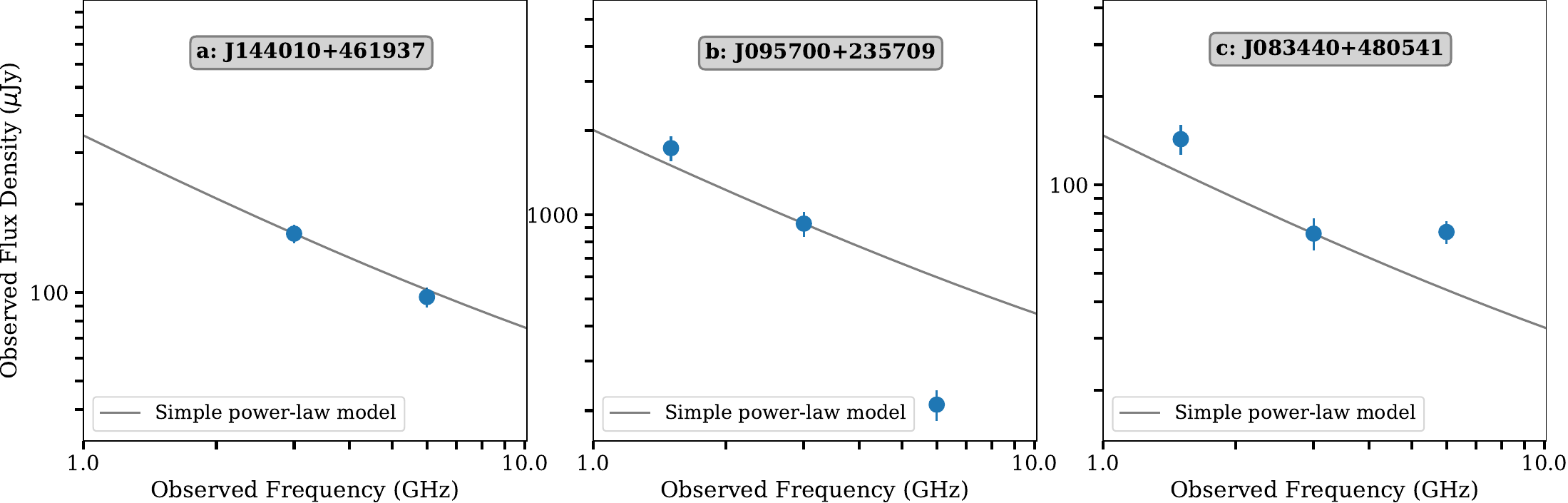}}
  \caption{Observed radio continuum SEDs at three frequencies: 1.5, 3, and 6 GHz for three sources in our sample. The error bars represent $1\sigma$ errors on the flux measurement. The gray line in each panel is an overlay (not a fit) for a simple radio-SED (Eq. \ref{eq: simple sed model}) normalized to the 3 GHz observed flux density with $\fth=0.1$ at 1 GHz and $\alphanth=-0.8$ commonly seen in star-forming galaxies. The left panel shows an example of a radio-SED which is close to the standard radio-SED (Eq. \ref{eq: simple sed model}). { The middle panel shows an example that has a steep spectrum above 3 GHz and is close to the standard radio-SED below 3 GHz.} 
  The right panel shows the case of a broken SED. }
  \label{fig: radio sed}
\end{figure*}

The RC in the simplest case is modeled by a single power law given by:\footnote{In the following, we use this sign convention to define the spectral index. Note:\ some references use the opposite sign convention.}$
    S_{\nu} = A \nu^{\alpha},$ where $\alpha$ is the power-law index, also termed the total spectral index, $\nu$ is the frequency, and $A$ is a proportionality constant. At frequencies below 30 GHz, the radio-SED for normal star-forming galaxies can be modeled as a combination of i) free-free thermal emission from star-forming regions having a relatively flat spectral index of $-0.1$ and ii) non-thermal emission having a steeper spectral index (\alphanth) arising due to synchrotron emission from relativistic electrons \citep[see e.g.,][]{condon1992}. Thus, the RC SED in such a case can be written with respect to a reference frequency $\nu_\circ$ following \citet{tabatabaei2017}, as:

\begin{equation}\label{eq: simple sed model}
    S_{\nu} = S_{\nu}^{\mathrm{th}} + S_{\nu}^{\mathrm{nth}} 
    = A_1\left(\frac{\nu}{\nu_\circ}\right)^{-0.1} + A_2 \left(\frac{\nu}{\nu_\circ}\right)^{\alpha_\mathrm{nth}}.
\end{equation} 

\noindent We can then define the thermal fraction at the reference frequency $\fth{} = S_{\nu_\circ}^{\mathrm{th}}/S_{\nu_\circ}$.

Under the assumption that the thermal emission is optically thin and no escape of LyC photons, $S_{\nu}^{\mathrm{th}}$ is directly related to the intrinsic ionizing photon rate and, thus, to the SFR and the electron temperature \citep[see e.g., Eqs. 10 \& 11 in ][]{murphy2011}. For normal star-forming galaxies, the non-thermal emission is dominated by CRs which were accelerated by supernova and diffuse in the ISM. Thus, the non-thermal emission is directly related to the supernova rate and \alphanth \citep[see Eq. 13 in ][]{murphy2011}. The relative contribution of the non-thermal component in normal star-forming galaxies increases as we go to lower frequencies (30 GHz and below). 
We show the radio-SED, for a normal star-forming galaxy with $\fth=0.1$ (10\%) at 1 GHz and $\alphanth=-0.8$ in Fig. \ref{fig: radio sed cartoon} as a dashed blue line. The dotted blue line shows the corresponding thermal emission component. The combined effect of the steep non-thermal and flat thermal component leads to an apparent spectral index between C- and S-bands (\alphacs{}) equal to -0.64. In the following, we refer to such a radio-SED and its apparent spectral index between C- and S-bands as standard and/or canonical, representing normal star-forming galaxies. We  use this reference value of \alphacs{} to compare with the spectral index of LzLCS sources.  

For young galaxies (ages $< 10$ Myr), it is possible that the \fth{} is higher than 0.1 due to a lack of enough supernovae and/or flat CR spectrum (>-0.5)  and thus closer to that found in SNRs \cite[e.g.,][]{Hunt04, Hunt05}. Moreover, in dense starbursts as discussed before, FFA has an effect on both the thermal and non-thermal emission and leads to a turnover in the radio spectra below a characteristic frequency \citep{condon1992}. This is termed the turnover frequency (\nuturn{}) and is the frequency at which the free-free optical depth (\odepth{}) reaches unity. The FFA \odepth{} is proportional to the emission \revtext{measure} (EM), \odepth{}$\propto \mathrm{n}_\mathrm{e}^2 \mathrm{L}$ (pc cm$^{-6}$), where $\mathrm{n}_\mathrm{e}$ is the electron number density and L is the path length along the line of sight. Thus \nuturn{} is also directly related to EM and increases with increasing EM. Following \citet{Clemens10} and \citet{Galvin18}, we can parameterize \odepth{} using \nuturn{}, such that \odepth{} $=(\nu / \nu_\mathrm{t})^{-2.1}$. We can thus modify the simple radio-SED from Eq. \ref{eq: simple sed model} under FFA in the simplest case as:
\begin{equation}
    S_\nu = (1 - e^{-\tau_\nu})\left [  A_1 + A_2  \left( \frac{\nu}{\nu_\circ} \right )^{0.1 + \alpha_\mathrm{nth}} \right ] \left( \frac{\nu}{\nu_\circ} \right )^{2}. \label{eq: FFA model}
\end{equation} 
In Fig \ref{fig: radio sed cartoon}, we show the effect of FFA with \nuturn{} at 1 GHz as orange line. In more complex scenarios, there can be multiple FFA components leading to even more complex features in the radio-SED \citep[e.g.,][]{Clemens10, Galvin18}.

Loss of CRs particularly due to synchrotron and inverse-Compton (IC) mechanisms can lead to curvatures in the spectrum towards high frequencies \citep{klein2018}. We show such a curved spectrum in Fig. \ref{fig: radio sed cartoon} with the red line. Finally, in the scenario of a single injection of CRs there can be a sharp cutoff at higher frequencies in the CR energy spectrum leading to a steepening at frequencies above a characteristic break frequency \citep[cutoff model in ][]{klein2018}. We show such a spectrum with a cuttoff with the green line in Fig. \ref{fig: radio sed cartoon}. We refer to \citet{klein2018} for a detailed discussion on various modifications to the standard radio-SED. The effect of steepening of the spectrum due to these mechanisms will show an effect on the spectral index in the frequency range explored in this work between (2-8 GHz), although we would need high-frequency (up to 30 GHz) observations to distinguish between different scenarios.

Fig. \ref{fig: radio sed} shows the observed RC SED for three LzLCS sources. In all three panels, the blue dots show the observed flux densities, the gray line corresponds to a simple radio-SED Eq. \ref{eq: simple sed model} normalized to the 3 GHz observed flux density with $\fth=0.1$ at 1 GHz and $\alphanth=-0.8$ which is typical of a star-forming galaxy. In the left panel is the radio-SED for J144010+461937 which has a radio-SED close to a star-forming galaxy. In Appendix \ref{appendix sec: All SEDs} we can see that several sources in our sample show similar SEDs.

 The middle panel shows the radio-SED for J095700+235709 where we see a steepening above 3 GHz. Several sources shown in Appendix \ref{appendix sec: All SEDs} are very steep at GHz frequencies. Moreover, the radio-SED of J082652+182052 and J091113+183108 shown in Appendix \ref{appendix sec: All SEDs} show evidence for FFA based on the 3-$\sigma$ upper limits on the L-band flux densities.
The right panel of Fig. \ref{fig: radio sed} exhibits the radio-SED of J083440+480541 which is an example of a broken SED above and below 3 GHz. J081409+211459 also shows a similar feature (Appendix \ref{appendix sec: All SEDs}).  Such a SED can arise due to a mix of multiple star-forming regions, one with an FFA component at high ($\sim$ 3 GHz), and the other with a steep spectrum. Multiple ULIRGs in the \citet{Clemens10} sample show such a radio-SED. Interestingly, we also find that these two sources are outliers when compared to their LyC properties which we discuss in Sec. \ref{sec: discussion}. Fig. \ref{appendix sec: All SEDs} shows the radio-SED for our entire sample of C- and S-band detections. Modelling the radio-SED to derive the relevant physical parameters has several free parameters and requires a flux density measurement across a large frequency range in order to remove the various degeneracies in parameter estimation \citep[see e.g.,][]{Galvin18}. Future observations of the radio-SED at lower frequencies (< 1 GHz, e.g., with the Giant Meterewave Radio Telescope and Low-Frequency Array) should help in constraining some of these physical mechanisms.

 \subsection{Total spectral index distribution}
\label{sec: spec index hist}

\begin{figure}
  \resizebox{\hsize}{!}{\includegraphics{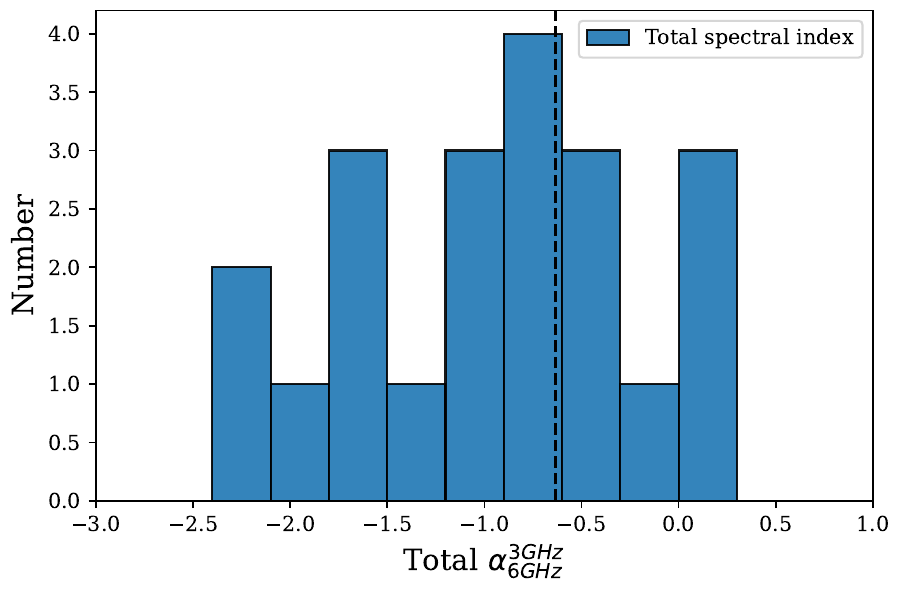}}
  \caption{Distribution of the total spectral index of LzLCS sources which spans a wide range from flat to steep. The dashed line shows the value of \alphacs{} expected in normal star-forming galaxies with a $\fth=0.1$ at 1 GHz and $\alphanth=-0.8$, \alphacs{} = -0.64. }
  \label{fig: spec index hist}
\end{figure}

We calculated the total spectral indices using the slope of the observed flux densities at 6 and 3 GHz (\alphacs) for our sources. We propagated the 1-$\sigma$ flux uncertainties on the flux measurements to estimate the uncertainties on the spectral {indices}. For sources with non-detections in only one of the S- and C-band, we estimated the corresponding lower/upper limit on the spectral indices using the $3\sigma$ upper limit on the flux density. In Table \ref{table: radio properties} we present the \alphacs{} values for our sample. From our sample of 24 C- and S-band detections, we could not derive reliable \alphacs{} for three sources due to unreliable flux density measurements. Here, the fluxes are unreliable due to the presence of strong sidelobes arising from very strong sources in the field. We use the rest of the sample of 21 sources with reliable \alphacs{} for all further analysis involving spectra indices. Overall in this sample 13 sources show steep (\alphacs{}$ < -0.64$) and 8 show flat radio spectra (\alphacs{}$ \geq -0.64$) between 2-8 GHz. For 7 of these sources we have L-band observations. Of them, 3 sources show a turnover at frequencies below 3 GHz, which is evidence for a FFA component between 1-3 GHz. And 2 sources show a broken radio-SED where the spectral {indices} between 1-3 GHz (L- and S-bands) is steeper than between 3-8 GHz (C- and S-bands). The remaining one source shows a typical radio-SED of a normal star-forming galaxy.

In Fig. \ref{fig: spec index hist}, we show the distribution of \alphacs{} in the blue histogram. { The mean \alphacs{} for our sample is $-0.92$ with a scatter of 0.71.} The total spectral index also varies in a very broad range from $-2.18$ to 0.22. 
On the contrary, the mean total spectral index of normal star-forming galaxies from the KINGFISHER galaxies was found to be $-0.79$  with a scatter of 0.15 \citep{tabatabaei2017} {, albiet in a wider frequency range of 1-10 GHz}. Thus LzLCS galaxies overall have a steeper mean radio spectra with a larger scatter than normal star-forming galaxies.

\begin{figure*}[ht]
  \resizebox{\hsize}{!}{\includegraphics{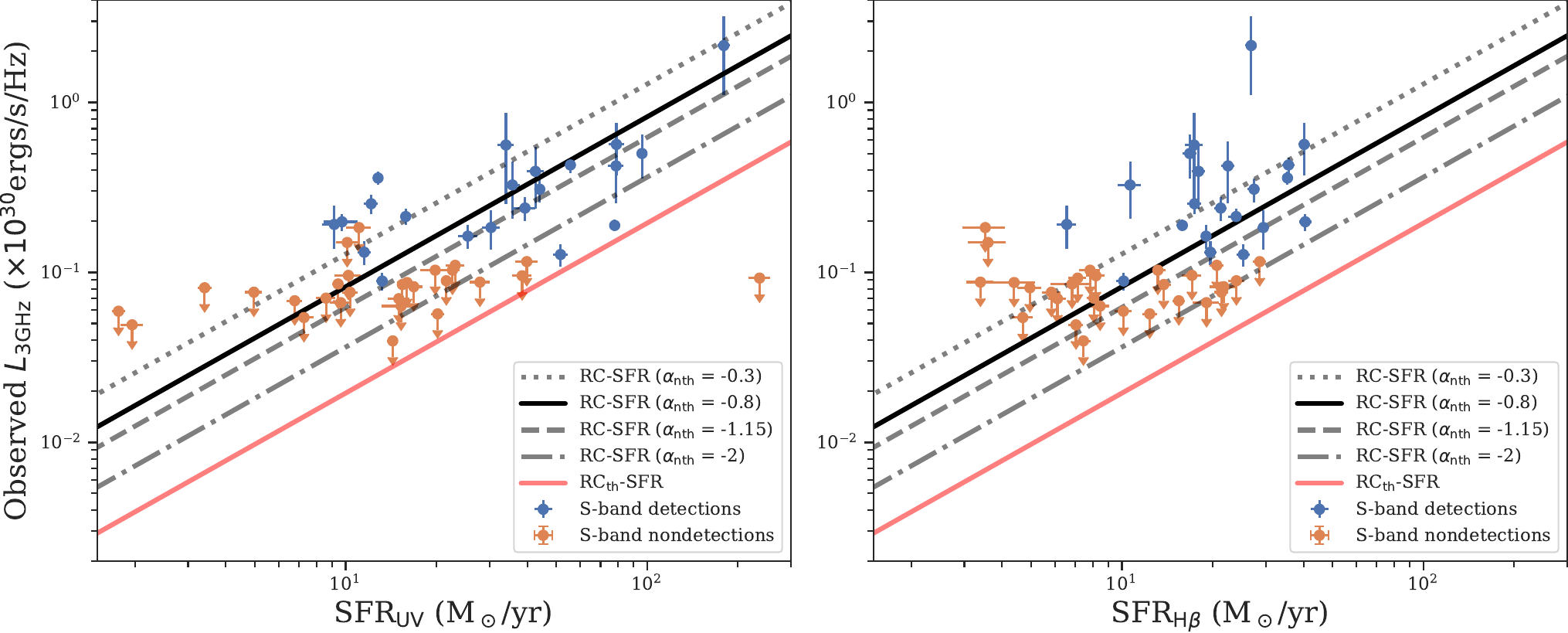}}
\resizebox{\hsize}{!}{\includegraphics{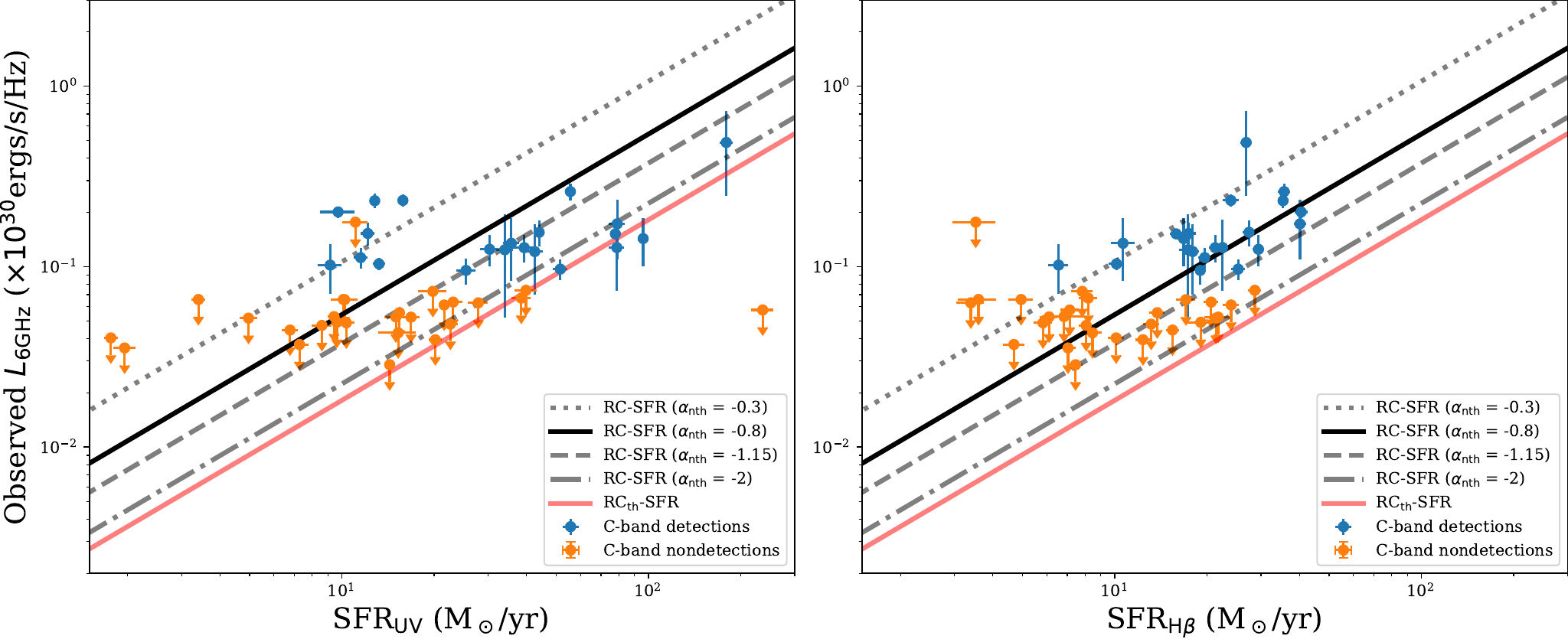}}
  \caption{{LzLCS sources on the RC-SFR relation at 3GHz top panels and 6GHz in the bottom panels. The left (right) panel shows the UV-based (H$\beta$-based) SFRs for the LzLCS sources. In all panels, the blue points represent detections in the corresponding band and the orange points represent the 3$\sigma$ upper limits for non-detections. The solid black line is plotted for the standard RC-SFR relation with $\alphanth{} = -0.8$. The dotted, dashed and dotted dashed lines are for different values of \alphanth{} from flat ($-0.3$) to very steep ($-2$), respectively. The red solid line represents the RC$_\mathrm{th}$-SFR relation from Eq. \ref{eq: RCth-SFR}. In all these cases, we have used a fixed value of T$_e = 10000$ K.}}
  \label{fig: RC SFR}
\end{figure*}

\subsection{The RC-SFR relation for LzLCS sources}
\label{sec: RC-SFR}

\begin{figure*}
  \resizebox{\hsize}{!}{\includegraphics{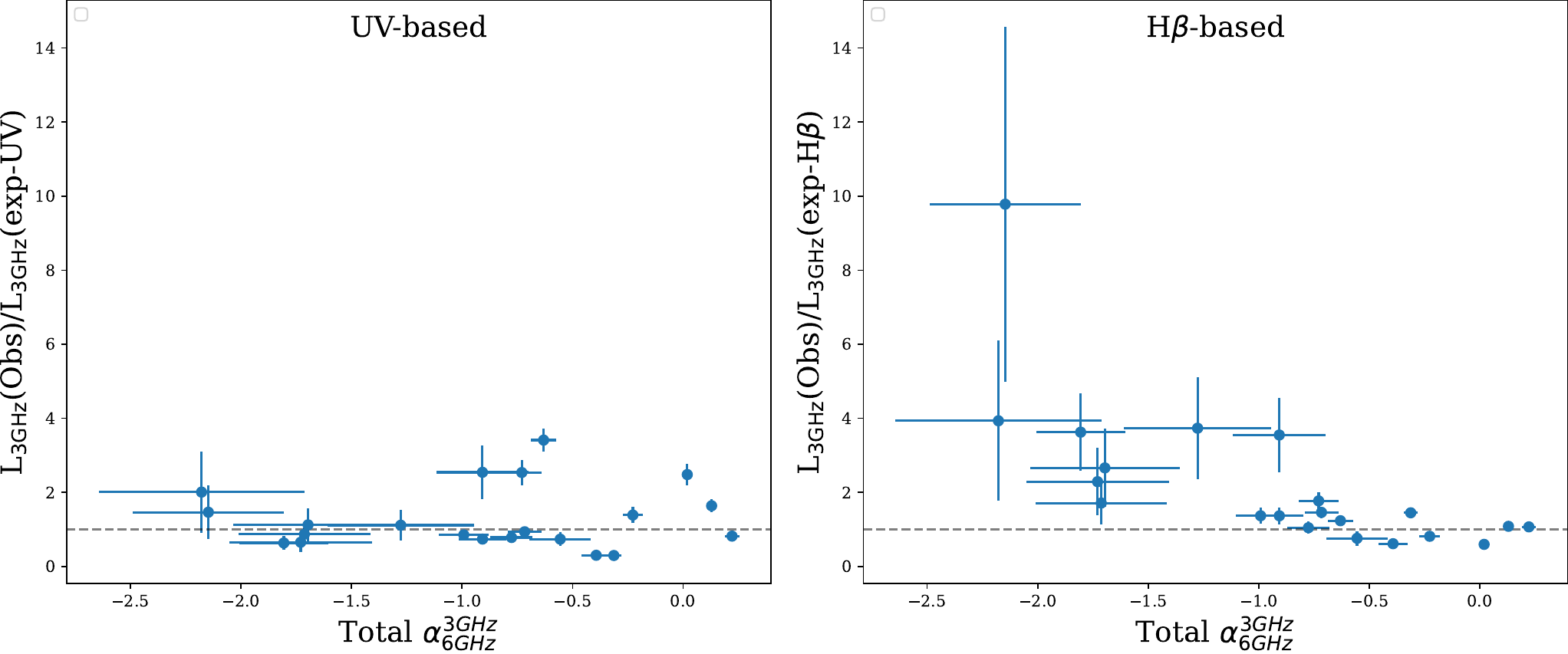}}
\resizebox{\hsize}{!}{\includegraphics{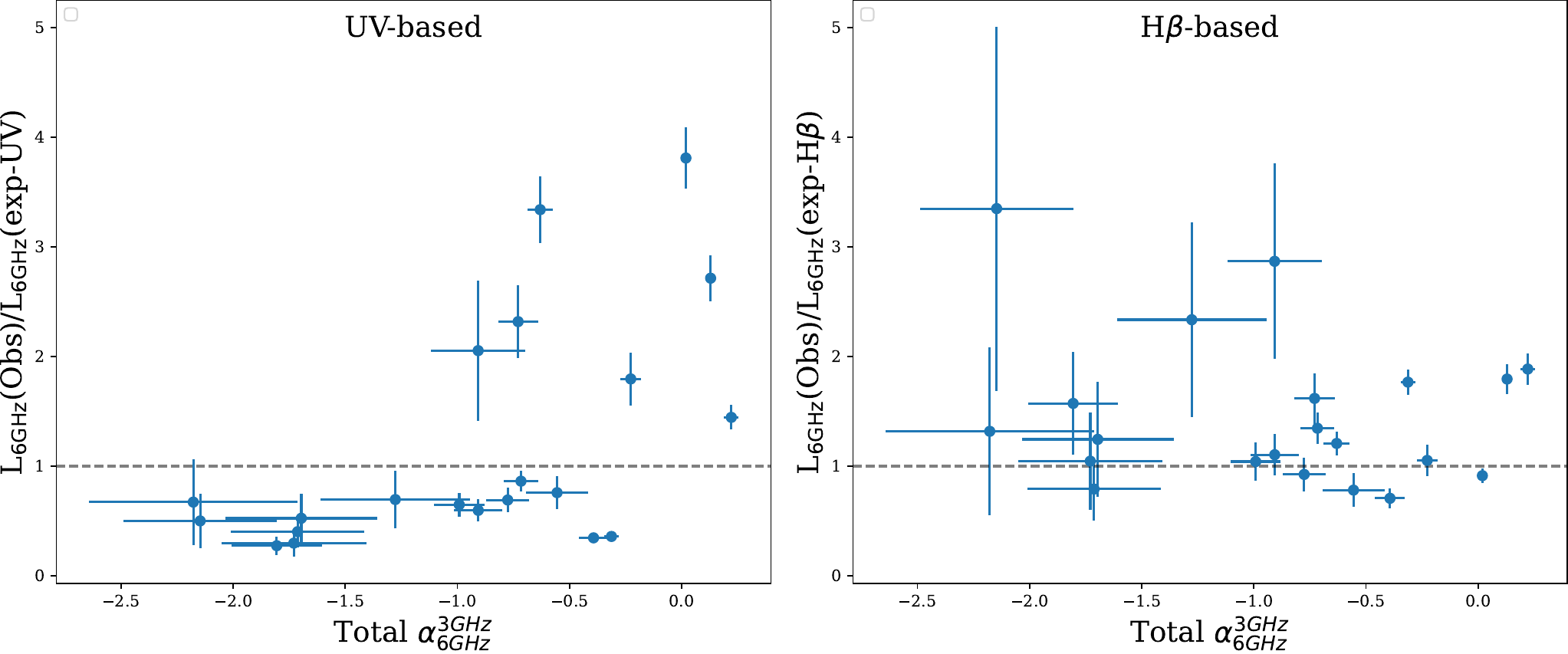}}
  \caption{{Ratio between the observed RC luminosity and that expected from the RC-SFR relation from a standard radio-SED with $\alphanth=-0.8$, $\fth=0.1$ at 1 GHz, and T$_\mathrm{e}$=10$^4$K, at 3 GHz and 6 GHz (top and bottom panels respectively) vs. the spectral index \alphacs{}. The left (right) panel shows the ratio with the expected RC for the UV-based (H$\beta$-based) SFR.  The dashed line shows a ratio of 1 for reference. Here, we have used only the S- and C-band detection sample.}}
  \label{fig: RC SFR offset}
\end{figure*}

The RC of normal star-forming galaxies at GHz frequencies can be used to determine their SFR \citep{Kennicutt12, murphy2011, tabatabaei2017}. In this section, we compare the observed RC of LzLCS galaxies to those predicted by such a calibration. The standard method used to calibrate the RC-SFR relation relies on the tight radio (at 1.4 GHz)-Far-infrared (FIR) correlation \citep{deJong85, Helou85}. However, since our observations are made at 3-6 GHz we need to scale to the 1.4 GHz luminosity  assuming a spectral index. Since LzLCS sources show a large variety in their radio-SED, such an assumption may not be valid across such a large frequency range.

An alternate approach as discussed in \citet{murphy2011} relates the thermal RC (\RCth{}) luminosity at a given frequency, $\nu$, to the SFR, as follows:
\begin{equation}
\label{eq: RCth-SFR}
    \mathrm{SFR}^{\mathrm{th}}_\nu =  4.6\times 10^{-28} \left ( \frac{T_e}{10^4~\mathrm{K}}\right )^{-0.45} \left (  \frac{\nu}{\mathrm{GHz}} \right )^{0.1} \left ( \frac{L^\mathrm{th}_{\nu}}{\mathrm{ergs/s/Hz}} \right ),
\end{equation}
where T$_e$ is the electron temperature. This is because both the SFR and \RCth{} emission are proportional to the ionizing photon rate in an optically thin plasma.
Moreover, the non-thermal RC (\RCnth{}) luminosity at a given frequency, $\nu$, can also be related to the SFR. This is because the SFR in a galaxy is related to the SN rate and so is the synchrotron emission at GHz frequencies. Therefore owing to a calibration between the SN rate and non-thermal luminosity \citep{ Condon-Yin90} we can derive a \RCnth{}-SFR calibration \citep[see Eq. 15 in ][]{murphy2011}. Here, it is assumed that the calibration between SN rate and non-thermal relation is the same as that observed in our Galaxy.  These two relations can then be combined to relate the total RC luminosity and the SFR as follows:
\begin{multline}
\label{eq: RCtotal-SFR}
    \mathrm{SFR}_\nu = 10^{-27} \left [ 2.18\left ( \frac{T_e}{10^4~\mathrm{K}}\right )^{0.45} \left (  \frac{\nu}{\mathrm{GHz}} \right )^{-0.1} \right .
   + \\ 
  \left . 15.1 \left( \frac{\nu}{\mathrm{GHz}} \right)^{\alpha_\mathrm{nth}} \right ]^{-1} \left ( \frac{L_{\nu}}{\mathrm{ergs/s/Hz}} \right ).
\end{multline}
Using this relation is more convenient than the radio-FIR correlation to compare galaxies with a non-standard \alphanth{} and thermal fraction.

{In Fig. \ref{fig: RC SFR}, we show the RC luminosity at 3 and 6 GHz for LzLCS sources against the extinction corrected UV-based SFR in the left panel and the H$\beta$-based SFR in the right panel. For converting the observed S-band (and C-band) flux density of LzLCS sources to the RC luminosity at 3 GHz (and 6 GHz) for the sample with S-band (and C-band) detections  we used the observed total \alphacs{}, and for the non-detections we assumed a total \alphacs$= -0.64$, as that of normal galaxies. As such there is a relatively weak dependence of the spectral index on the luminosities as the K-corrections are small since our sources are at a relatively low redshift of $\sim 0.3$. The blue points in both panels are for the C- (and S-band) detection samples and the orange points are the 3$\sigma$ upper limits for the sample with S-band (and C-band) non-detections. }

{In Fig. \ref{fig: RC SFR}, the solid black lines show the total RC-SFR calibration from Eq. \ref{eq: RCtotal-SFR} in all the panels. Here, we assumed a standard $\alphanth=-0.8$ and T$_e = 10000$ K. We also show how this relation changes purely if the \alphanth{} steepens to $-1.15$ and $-2$.
 Steep values of $\alphanth \sim -2.1$ are observed in nearby BCDs \citep{Klein91}, and this is the reason we chose this range.  We note that the relation has the same slope, but it systematically moves down as the \alphanth{} steepens. This happens because, for a fixed frequency, a steeper non-thermal spectrum has a lower contribution from the non-thermal emission. On the contrary, a flatter \alphanth{} moves the relation upwards. 
 In the case where there is no non-thermal emission namely, purely thermal radio spectra, we use the \RCth{}-SFR relation from Eq. \ref{eq: RCth-SFR}. The red solid lines in both panels show this limit. In principle, galaxies can also be below this line if the thermal emission is reduced due to FFA around 3 GHz. Such FFA at GHz frequencies is observed in compact starbursts with high EM \citep[e.g.,][]{Hunt04}. Most of the points in Fig. \ref{fig: RC SFR} are above the red line due to the contribution of non-thermal emission in LzLCS sources. We compare the RC luminosity with the SFR derived from two different tracers, UV-based and H$\beta$-based. 
Thermal radio emission is due to the Coulomb interaction of free thermal electrons in an ionized gas and H$\beta$ emission arises due to the recombination of electrons in the same region.
And the UV emission is simple starlight arising from massive OB stars. The timescale of {star-formation} traced by the H-$\beta$ line is of the order of 10 Myrs and is much shorter than the star-formation timescale traced by the UV emission which is $\sim100$ Myrs \citep{Kennicutt12}. }

{
Overall, we observe that although LzLCS sources show a correlation between their RC luminosity and SFR, there is a lot of scatter from the standard RC-SFR relation, with several sources much above and below this relation. It is important to note that several points are above the \RCth{}-SFR relation, thus suggesting that LzLCS sources have non-thermal emission which suggests past SN activity.}




\begin{figure}[htb]
  \resizebox{\hsize}{!}{\includegraphics{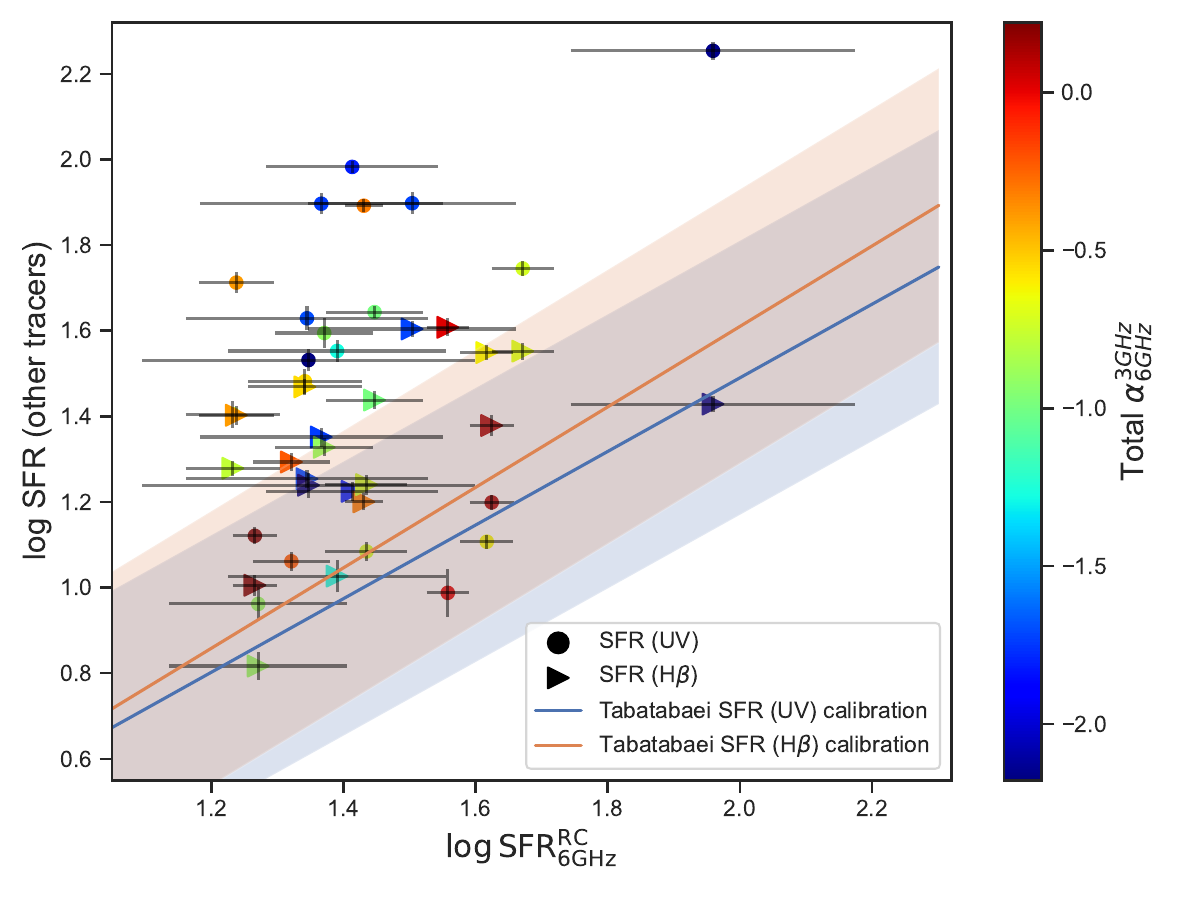}}
  \caption{ Comparison between \SFRRC{6 GHz} and \SFRUV{} (circles) and \SFRHbeta{} (triangles). Each of the points is color coded by their apparent \alphacs{}. We over plot the relation between \SFRRC{6 GHz} and other tracers from \citet{tabatabaei2017} in blue (\SFRUV{} calibration) and orange (\SFRHbeta{}  calibration). The blue and orange shaded region shows the scatter in these relations. }
  \label{fig: SFR RC comparison}
\end{figure}

\begin{figure}[htb]
  \resizebox{\hsize}{!}{\includegraphics{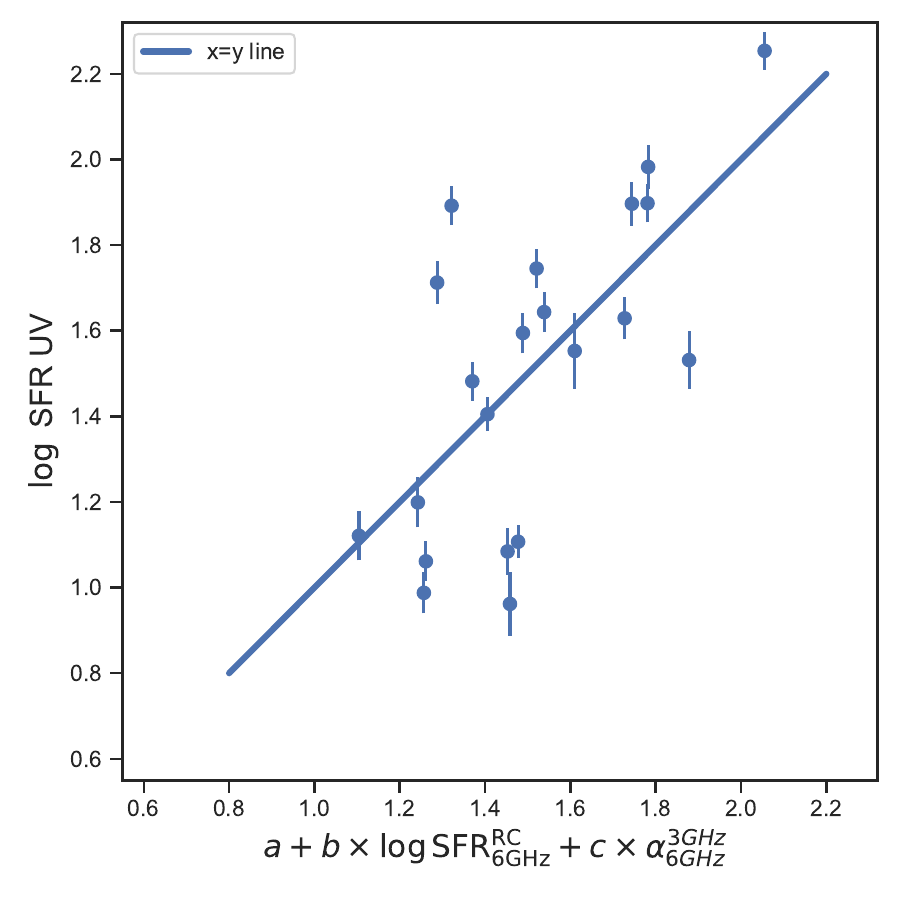}}
  \caption{ {Combined fit between the linear combination of $\log_{10}$ \SFRRC{6 GHz} and \alphacs{}, and $\log_{10}$ \SFRUV{}. The blue line shows the $x = y$ line. Note: the residual in the fit is 0.21, which is similar to the typical scatter between different SFR tracers from \citet{tabatabaei2017}.} 
  }\label{fig: SFR RC fit}
\end{figure}

\begin{figure*}[htb]
  \resizebox{\hsize}{!}{\includegraphics{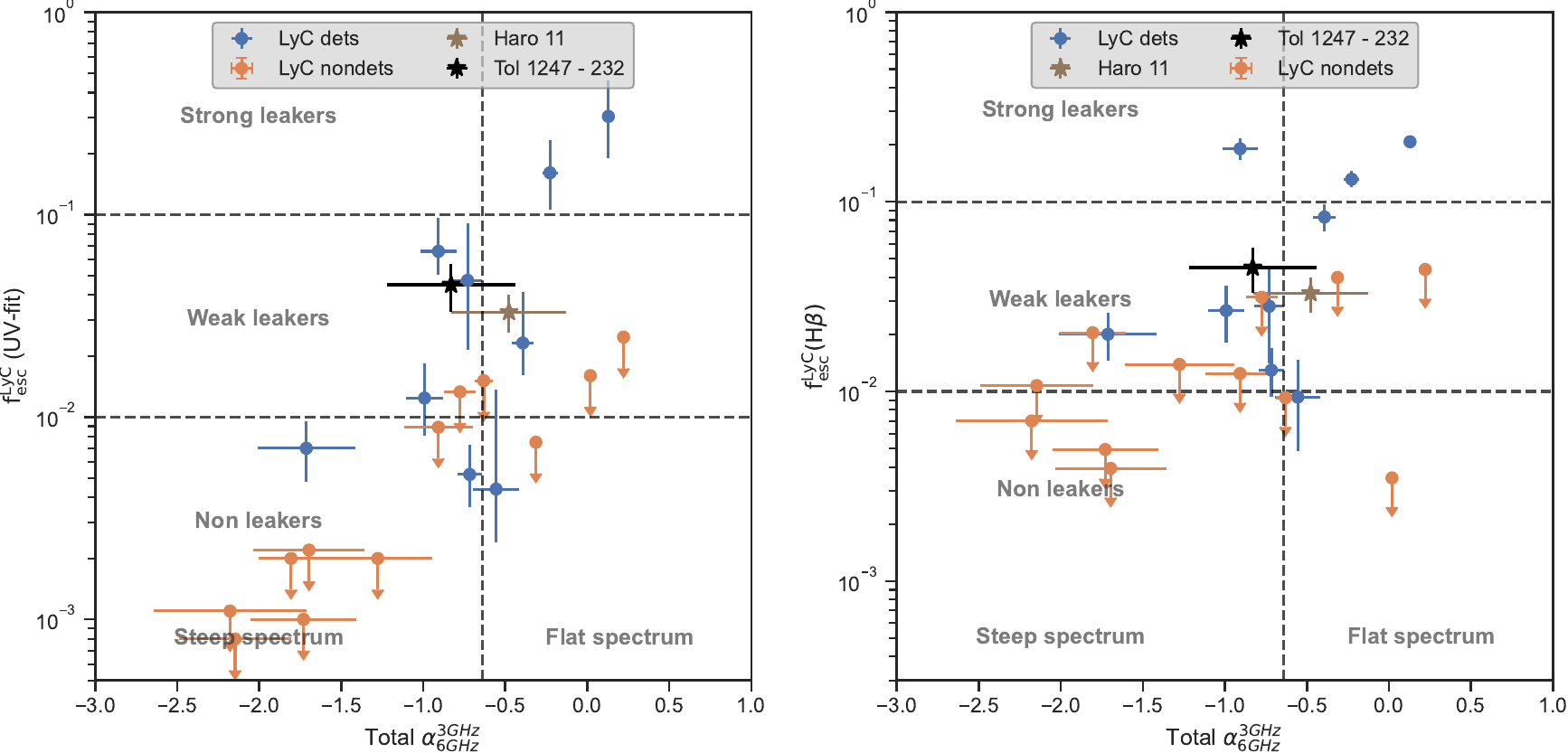}}
  \caption{{Dependence of \fesc{} on the total spectral index measured across C- (6 GHz) and S- (3 GHz) bands.} The left panel shows the dependence on \fesc{} measured using the UV-fit method. And the right panel shows the dependence on \fesc{} measured using the H$\beta$ method \cite{Flury22a}. The blue points in both panels show LyC detections and the orange points show LyC non-detections (1$\sigma$ upper limits). We also show two known nearby leakers from the literature, Haro 11 (brown star) and Tol 1247-232 (black star). In both panels the horizontal lines demarcate the regions for strong leakers (\fesc{} > 0.1), weak leakers ( 0.01 < \fesc{} < 0.1), and non-leakers (\fesc{} < 0.01). The vertical line shows the region above and below a spectral index of a standard radio-SED (\alphacs{}$= -0.64$) demarcating the region for flat and steep spectrum sources. See text for details. 
  }
  \label{fig: total spec index fesc}
\end{figure*}

We next investigate if the offset between the observed and expected RC luminosity can be explained by variations in the observed \alphacs{} of these sources. In Fig. \ref{fig: RC SFR offset}, we show the ratio between the observed RC luminosity and that expected from the RC-SFR relation based on Eq. \ref{eq: RCtotal-SFR} (termed as the ``SFR offset'' here on) with respect to the total \alphacs{} for only the C- and S-band detection sample. {To estimate the RC luminosities here we have used the observed spectral index values as done previously.} We study the RC-offset for both the S- and C-bands and for SFR measured using UV and H$\beta$ emission. We notice that the H$\beta$-based SFR offset systematically increases for steep spectrum sources (top right panel) for the S-band, whereas the flat spectrum sources are closer to that expected. The SFR offsets with UV-based SFRs at C-band are higher for flat spectrum sources (bottom left panel). Such a deviation can lead to values above the expected RC-SFR relation in Fig. \ref{fig: RC SFR}. The SFR offsets with UV-based SFRs for S-band (top left panel) and H$\beta$-based SFRs for C-band (bottom right panel) do not show any such offsets although they have a large scatter. We currently do not completely understand the origin of such systematic and large offsets. Overall this suggests that accurate SFR estimation using different tracers is difficult for LzLCS sources and particularly so with radio-based SFR tracers.

Finally, an important reason for the scatter in the RC-SFR relation can be that for galaxies underdoing starbursts like the LzLCS sources, the SFR traced by different methods are in
general not consistent due to the stochastic/bursty nature of star formation \citep[e.g.,][]{Izotov16c, Atek22}.
This inconsistency is also evident between the UV-based and H$\beta$-based
SFRs. Finally, for the leakers (weak and strong) the H$\beta$-based SFR can underestimate the true SFR, since not all the ionizing photons are absorbed by the star-forming regions. Most of the leakers in our sample are weak ($< 10$\%) and the two strong leakers have \fesc{} below $30$\%. Thus this effect is marginal compared to the large SFR offsets we observe. We also see that the RC is significantly below the RC-SFR
relation in several sources. 
This can happen due to a combination of a steep \alphanth{} or a relatively low non-thermal fraction. In a few sources, their RC luminosity is fairly close to that expected from the \RCth{}-SFR relation, and are thus possibly thermally dominated systems. Such a deficit has also been observed in a sample of green-peas \citep{chakraborti2012}, blueberries \citep{Sebastian19} and in \hii{} galaxies \citep{Rosa-Gonzalez07}. The effect of FFA on both thermal and non-thermal emissions can also be a further cause of such a deficit. A few sources are above the black solid line, where it is possible that their \alphanth{} is flatter than the canonical value. It is also possible that the calibration between the SN rate-SFR observed in our Galaxy is not valid for LzLCS sources.

\subsection{{Comparison between radio-based SFR with other SFR tracers} }

We compare the radio-based SFR derived at 6 GHz (\SFRRC{6 GHz}) with the \SFRUV{} and \SFRHbeta{}  for our LzLCS sample in Fig. \ref{fig: SFR RC comparison}. We use Eq. \ref{eq: RCtotal-SFR} to derive the \SFRRC{} where we assumed a \alphanth{} of $-0.8$. We find that the \SFRRC{} shows a systematic offset from the \SFRRC{6 GHz}-\SFRUV~(\SFRHbeta) relation for local normal star-forming galaxies from \citet{tabatabaei2017} similar to that observed in the bottom-left panel of Fig. \ref{fig: RC SFR}. The median value of this offset from \SFRUV{} (\SFRHbeta{}) {is $-0.59$ dex ($-0.26$ dex)}. The only difference between the SFR calibrations from \citet{tabatabaei2017} is that it uses H$\alpha$ based SFR as opposed to H$\beta$ line used in our work. {\citet{tabatabaei2017} found a slightly higher value of \alphanth{} of -0.97, in their KINGFISHER sample. Assuming this value to calculate the \SFRRC{} decreases the offsets by a small amount to $-0.52$ dex ($-0.18$ dex) for \SFRUV{} (\SFRHbeta{})}.

We also color-code the points in Fig. \ref{fig: SFR RC comparison} with \alphacs{} to study it dependence of the SFR offset. As discussed in Sec. \ref{sec: RC-SFR} we observe that the ratio of the observed and expected RC luminosity shows a systematic offset with \alphacs{}. We hence fit the logarithm of \SFRUV{} (\SFRHbeta)  with a linear combination of $\log_{10}$ \SFRRC{6 GHz} and \alphacs{} as follows:

\begin{equation} \label{eq: SFR RC fit}
\log_{10} \mathrm{SFR}_\mathrm{UV (H\beta)}  = \mathrm{a} + \mathrm{b} \times \log_{10} \mathrm{SFR}^\nu_{\mathrm{RC}} + c\times \alpha^{\mathrm{3GHz}}_\mathrm{6GHz}.
\end{equation}
{ The details on the fitting procedure can be found in Appendix \ref{appendix sec: SFR fit details}. We show this fit between $\log_{10} \mathrm{SFR}_\mathrm{UV}$ and $\log_{10}$ \SFRRC{6 GHz} and \alphacs{} in Fig. \ref{fig: SFR RC fit} where the residual is 0.21. This is comparable to the typical scatter found between different SFR tracers from \citet{tabatabaei2017}. We provide the fit parameters at different radio frequencies in Table \ref{table: SFR RC fit parameters}. In the future, with observations spanning a wide frequency range from 1-10 GHz, we can use the Mid RC (MRC)-SFR calibration suggested by \citep{tabatabaei2017} to further improve the \SFRRC{} calibration.}

  \begin{table*}[ht!]
      \caption[]{{\SFRRC{\nu} fit coefficients to various other SFR tracers.}}
         \label{table: SFR RC fit parameters}
     \centering

        \begin{tabular}{ccccccc}
        \hline \hline \\
X$_{1}$ & X$_{2}$ & Y & a & b & c & $\sigma$ \\ 
\hline \\
\SFRRC{6 GHz} &  \alphacs{}  & \SFRUV{}  & 0.79 $\pm$ 0.25     &  0.30 $\pm$ 0.18 &  -0.31 $\pm$ 0.03 &  0.21 \\ \\

\SFRRC{6 GHz}  &  \alphacs{}  & \SFRHbeta{}  &  0.69 $\pm$ 0.21 & 0.46 $\pm$ 0.15  &  0.02 $\pm$ 0.03 & 0.14  \\ \\

\SFRRC{3 GHz} &  \alphacs{}  &  \SFRUV{} & 0.82 $\pm$ 0.21  & 0.32 $\pm$ 0.17  & -0.22 $\pm$ 0.07  &  0.21 \\ \\                        


\SFRRC{3 GHz} &  \alphacs{}  & \SFRHbeta{}  & 0.77 $\pm$ 0.18  &  0.45 $\pm$ 0.15 & 0.15 $\pm$ 0.06  & 0.14 \\ \\
                         \hline \\
\end{tabular}
\tablefoot{{The linear fit is performed in logarithmic scales as: $\log Y = a + b\times \log X_{1} + c\times X_{2}$.} Here,  $X_2$ is \alphacs{} in all cases. And $\sigma$ represents the median value of the residual from the corresponding fit.}     
\end{table*}


\begin{figure*}[htb]
  \resizebox{\hsize}{!}{\includegraphics{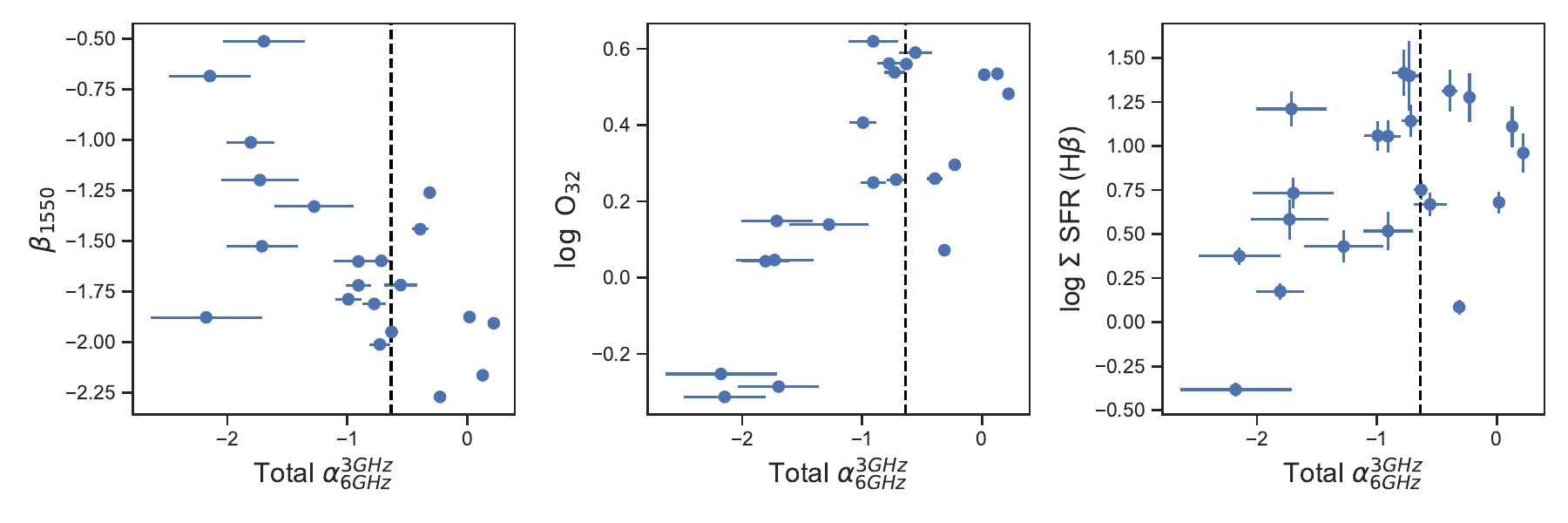}}
  \caption{ { Dependence of \betatwo{},  \Oratio{} and \sfrdensity{} on the spectral index \alphacs{} for the LzLCS sample.} The vertical line demarcates the flat (\alphacs{}$ > -0.64$) and steep spectrum (\alphacs{}$< -0.64$) sources. }
  \label{fig: phys properties alpha}
\end{figure*}

\subsection{Relations between LyC escape and RC emission}

\label{sec: specindex vs fesc}

{In this section, we for the first time study the dependence of the Lyman continuum escape fraction, \fesc, on the total spectral index and thermal fraction empirically.} Figure \ref{fig: total spec index fesc} shows the dependence of \fesc{} on the total spectral index measured in the two bands, C- and S. For this and all further analysis we use only the sources with reliable \alphacs{} for our analysis. The sources with LyC detections are shown in blue points and the non-detections are shown as upper limits (1-$\sigma$) in orange points.

The left and right panels shows the radio spectral index dependence {of} \fesc{} measured using two different approaches, named the ``UV-fit'' and ``H$\beta$'' methods \citep[see][for details]{Flury22a}. In both methods, the escape fraction is calculated as the ratio between the "observed" (escaping) and the "intrinsic" (produced) ionizing fluxes, both measured at a narrow wavelength interval just below the Lyman edge (912\AA). {Both former approaches differ according to the way the intrinsic ionizing flux is estimated.} In the UV-fit method, the intrinsic LyC flux was predicted via detailed SED-fitting to non-ionizing part of the COS/G140L spectra of LzLCS galaxies \citep[see][]{Saldana-Lopez22}. In the H$\beta$ method, the intrinsic LyC flux density was estimated from the measured equivalent-width of the H$\beta$ line from SDSS spectra, as a tracer of the age of the most massive stellar populations \citep[see also][]{Izotov16b}. In instantaneous-burst models \citep[][used consistently for both former approaches]{Leitherer1999}, the age of the burst is inherently related to the production of LyC photons.


The horizontal dashed lines show the region for strong leakers (\fesc{} > 0.1), weak leakers ( 0.01 < \fesc{} < 0.1), and non-leakers (\fesc{} < 0.01). The vertical dashed line shows the value of \alphacs{} for a standard radio-SED from Eq. \ref{eq: simple sed model} with a $\alphanth=-0.8$ and $\fth= 0.1$ at 1 GHz. This line splits the figure into regions with flat spectrum (\alphacs{} $> -0.64$) and steep spectrum sources (\alphacs{} $< -0.64$). We also show two known nearby weak leakers, Haro 11 and Tol 1247-232 on this plot. We used the literature value of \fesc{} from \citet{Leitet11} and \citet{Leitherer16} for Haro 11 and Tol 1247-232 respectively. The radio spectral index for Haro 11 was measured between 3.0-8.46 GHz. We used \citet{Schmitt06} for the 8.46 GHz flux density and the Very Large Array Sky Survey (VLASS) catalogue \citep{Gordon21} for the 3.0 GHz flux density. The radio spectral index for Tol 1247-232 was measured between 5-8 GHz using VLA observations from \citet{Rosa-Gonzalez07}. The radio spectral index is measured in a very similar range frequency range as our C- and S-bands, enabling us to use them for comparison purposes.

It is interesting to find
a positive correlation between \fesc{}, particularly measured using the UV-fit and \alphacs{}. The correlation appears to be marginally weak for the \fesc{} measured using H$\beta$, however, that seems to be largely driven by the difference in the upper limits on the LyC non-detections between the two methods. This indicates that strong leakers show overall flatter radio spectra, and non-leakers have steeper \alphacs{}, while weak leakers typically have \alphacs{} close to the canonical value (\alphacs{} $= -0.64$). Interestingly, Haro 11 and Tol 1247-232 are also within our \alphacs{}-\fesc{} relation. We also find two outliers (J155945+403325 and J083440+480541) from the \alphacs{}-\fesc{} relation (in both the plots) and J081409+211459 which is clearly an outlier in the left-panel of Fig. \ref{fig: total spec index fesc} and possibly in the right panel. These sources show a flat spectrum but much lower values of \fesc{} than other LzLCS sources. Interestingly, J083440+480541 and J081409+211459 have L-band observations and show a steeper spectrum at low frequencies. However, in other sources with L-band observations we see either a lower L-band flux density or that consistent with a standard radio-SED. Finally in Table \ref{table: kendall tau} we quantify these correlations using the Kendal $\tau$ correlation coeffiecient. Here we take into account the upper limits on the \fesc{} values following \citet{Akritas96} which is the procedure following in \citet{Flury22b}. We find that \alphacs{} correlates with both \fesc{} UV-fit and H$\beta$ method (strong $\tau$ values) particularly, after removing the two outliers. However the significance of this correlation is less compared to other diagnostics discussed in \citet{Flury22b}. It is important to note that our current sample is quite small compared to the full sample studied in \citet{Flury22b} and also lacks sources at high LyC escape. In the future, we aim to increase the sample size with radio observations of LyC leakers, and  particularly with L-band data. This should give us a better picture of the dependance of \fesc{} and \alphacs{} and on the properties of outliers.







\subsection{Relation between \betatwo{}, \Oratio{}, and \sfrdensity{} {and} the RC emission}
\label{sec: phys properties alpha}

The escape of LyC photons from galaxies is known to be correlated with several other physical parameters from the literature \citep[e.g.,][]{Izotov21b, Flury22b, Saldana-Lopez22, Chisholm22, Marques-Chaves22b, Xu23}, particularly the UV slope \betatwo{}, the \Oratio{} emission line ratio, and SFR surface density \sfrdensity{}.
In particular, galaxies with high \fesc{} are expected to have steeper values of \betatwo{}, higher \Oratio{}, and a higher \sfrdensity{} and vice versa. Thus the correlation between \alphacs{} and \fesc{} observed here leads us to explore if the \alphacs{} of LzLCS sources also depends on these parameters with similar trends. Moreover, the LzLCS sample is also selected based on a range of $\beta$, \Oratio{}, and \sfrdensity{}. { In Fig.~\ref{fig: phys properties alpha}, we show the dependence of \betatwo{}, \Oratio{}, and \sfrdensity{} on \alphacs{}.} We see that \alphacs{} and \betatwo{} are generally anti-correlated, so the \betatwo{} slope steepens as \alphacs{} flattens. The presence of flatter \betatwo{} slopes (i.e.,\ redder UV spectra) in sources with steep \alphacs{} can be associated with the declining phase of the starburst (see Sec. \ref{sec: non-leakers discussion} for more discussion) and also to higher dust attenuation \citep[e.g.,][]{Saldana-Lopez22, Chisholm22}. We also notice that \alphacs{} overall flattens with increasing \Oratio{} and \sfrdensity{} which is both in line with that expected with their dependence on \fesc{}. However, it is difficult to associate a direct causal relation between \alphacs{} and these parameters as \fesc{} depends on multiple physical parameters \citep[e.g.,][]{Flury22b}. We report the Kendall $\tau$ values between these quantities in Table \ref{table: kendall tau}. We find a strong correlation between these quantities but with a low significance (low $p-$values).  We discuss the possible physical reasons driving a correlation between \alphacs{} and \fesc{} in the next section.

\begin{table}[ht!]
\centering
\caption{{Kendall $\tau$ correlation coefficients and the $p-$values between the radio spectral index (\alphacs{}) versus \fesc{} and different physical properties of LzLCS sample.} }\label{table: kendall tau}

\begin{tabular}{ccc}
\hline\hline \\
\multicolumn{1}{c}{} & \multicolumn{2}{c}{\alphacs{}} \\
\midrule
 & $\tau$ & $p_{val}$ \\ 
\cmidrule{2-3} 
\fesc{}(UV-fit)$^*$\ & $+0.395 (0.285)$ & $1.220 \times 10^{-2}$ ($5.723 \times 10^{-2}$) \\
\fesc{}(H$\beta$)$^*$\ & $+0.381 (0.261)$ & $1.57 \times 10^{-2}$ ($8.132 \times 10^{-2}$) \\
\betatwo{}\ & $-0.419$ & $7.876 \times 10^{-3}$ \\
$\log_{10}$\Oratio{}\ & $+0.448$ & $4.532 \times 10^{-3}$ \\
$\log_{10}$\sfrdensity{}(H$\beta$)\ & $+0.286$ & $7.001 \times 10^{-2}$ \\
$\log_{10}$\mstar{} & $-0.410$ & $9.406 \times 10^{-3}$ \\
\bottomrule
\end{tabular}
\tablefoot{$^{*}$ Here, we take into account the upper limits on \fesc{} (following \citet{Akritas96}) with the same approach as described in \cite{Flury22b}. The $\tau$ values are estimated after removing the two outliers and the values in the bracket are including the outliers. The typical uncertainty in $\tau$ is $\pm$0.11.}
\end{table}

\section{Discussion}
\label{sec: discussion}

\subsection{Estimated thermal fraction of LzLCS sources}
\label{sec: thermal fraction hist}

\begin{figure}[ht]
  \resizebox{\hsize}{!}{\includegraphics{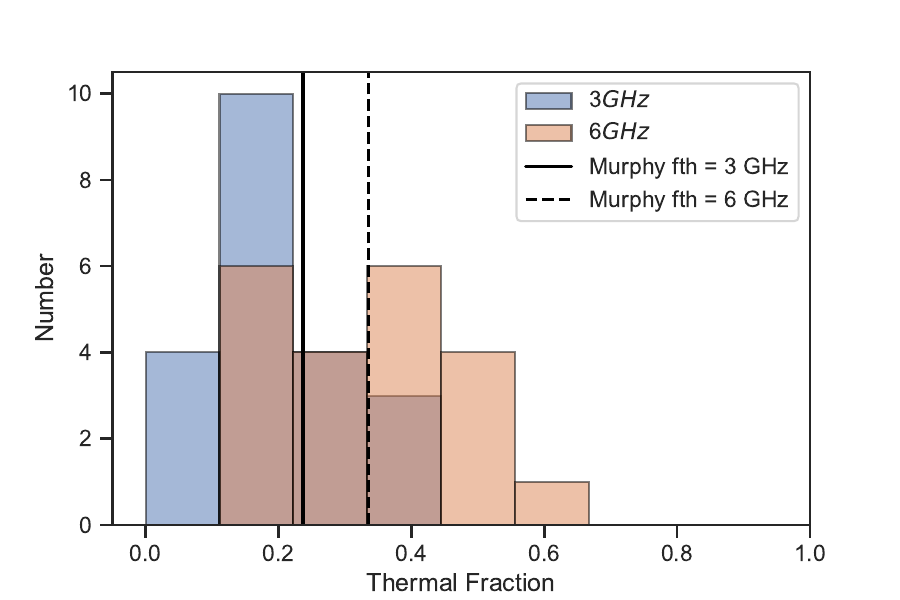}}
  \caption{Distribution of the estimated thermal fraction for LzLCS sources. The blue (orange) histograms show the thermal fraction at 3 GHz (6 GHz). The  solid (dashed) line shows the expected thermal fraction at 3 GHz (6 GHz) for normal star-forming galaxies with a $\fth=0.1$ at 1 GHz and $\alphanth=-0.8$.}
  \label{fig: thermal fraction hist}
\end{figure}

\begin{figure}
  \resizebox{\hsize}{!}{\includegraphics{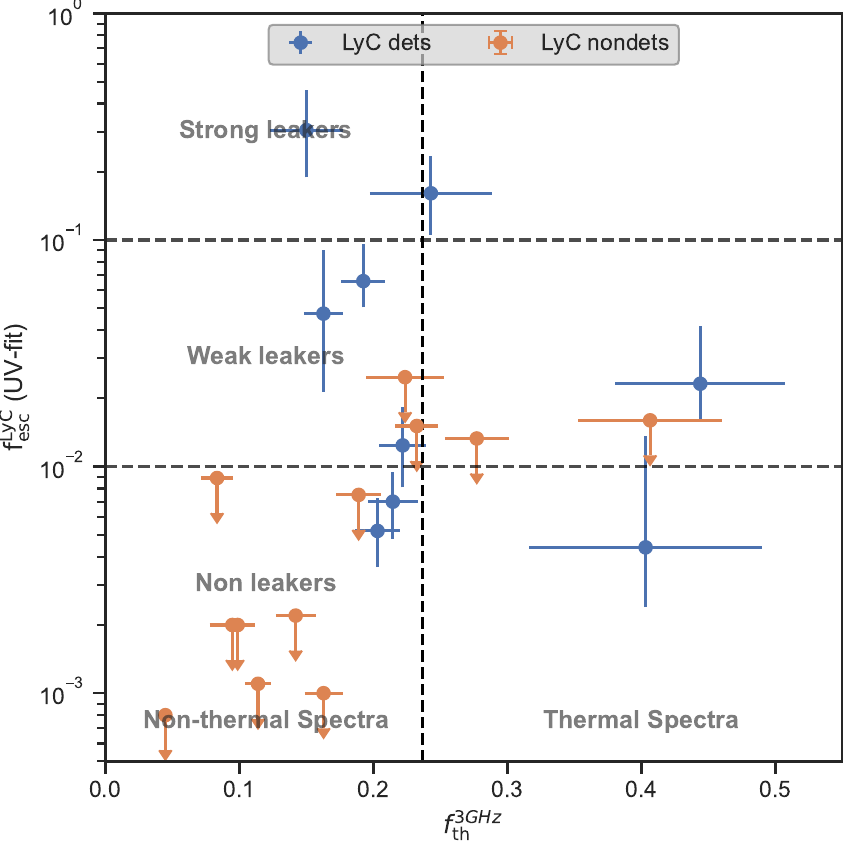}}
  \caption{ { Dependence of \fesc{} measured using the UV-fit on the estimated thermal fraction at 3 GHz.} The horizontal dashed lines demarcate the region for strong, weak and non-leakers. The vertical dashed line shows the expected thermal fraction at 3 GHz for normal star-forming galaxies with a $\fth=0.1$ at 1 GHz and $\alphanth=-0.8$. }
  
  \label{fig: thermal frac fesc}
\end{figure}

We estimated the thermal flux density at two different observed frequencies, 3 GHz and 6 GHz, using Eq. \ref{eq: RCth-SFR}. We use the SFR (derived from the dust-corrected H$\beta$ line intensity) and electron temperature derived using [OIII] $\lambda\lambda 4363; 4959,5007$ \citep[see][for details]{Flury22a}. {We further correct this thermal flux density by a factor of (1 - \fesc{}) to account for the escaping LyC photons. We use this thermal flux density and the corresponding observed flux density at 3 GHz (6 GHz) to derive the thermal fraction at these two frequencies.}

{The mean value of the thermal fraction at 3 GHz (6 GHz) is  $0.21 \pm 0.11 ~(0.35 \pm 0.14)$.} {The mean thermal fraction at 6 GHz for nearby star-forming galaxies from the KINGFISHER sample was found to be $ 0.23 \pm 0.13$ \citep{tabatabaei2017} which is slightly higher although within the uncertainties.} Moreover, the range of the thermal fractions of our sample is similar to that of the KINGFISHER sample. Fig. \ref{fig: thermal fraction hist} shows the distribution of the thermal fraction at 3 GHz in blue and 6 GHz in the orange histogram. The fact that the radio emission from these galaxies is not dominated by thermal emission provides evidence for a non-zero contribution from synchrotron emission in these galaxies. Interestingly, this provides the first direct evidence of SN activity in these galaxies. 

Next, we investigate if SN activity could play a role in facilitating LyC escape as expected from various simulations \citep[e.g.,][]{Trebitsch17, Ma20}. In Fig. \ref{fig: thermal frac fesc} we study the dependence of \fesc{} on the thermal fraction which informs us about increased SN activity or the lack of it. We do not observe any correlation, rather most of our sources including the strongest leakers have thermal fractions close to that expected for normal star-forming galaxies (vertical dashed line). Most of the non-leakers instead are non-thermal dominated thus suggesting older episodes of star formation. A similar lack of dependence is observed between the thermal fraction derived at 6 GHz and \fesc{}. 

{A few sources in our sample are very compact starbursts where it is common to have a significant FFA component. Therefore, it is possible that the true thermal fractions can be even higher than our estimates. A complete radio-SED fitting, particularly with even higher frequency data, is required to accurately constrain the thermal fractions.}

\subsection{Considering why strong leakers have a flat radio spectrum }

\begin{figure}
  \resizebox{\hsize}{!}{\includegraphics{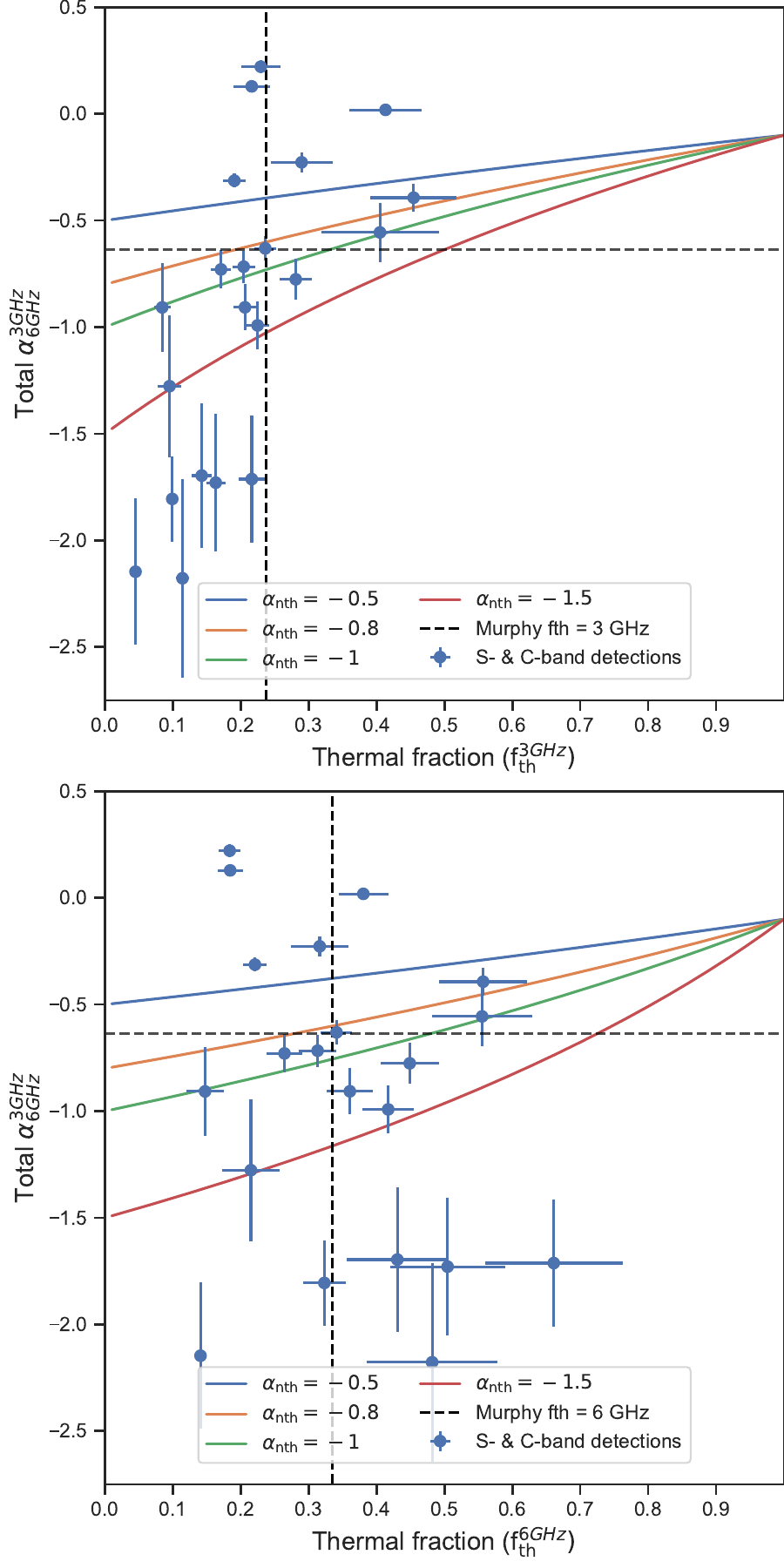}}
  \caption{Effect of increasing \fth{} at a reference frequency of 3 GHz (top panel) and 6 GHz (bottom panel) on the observed spectral index \alphacs{}. Here we consider the simplest radio-SED composed of a thermal and non-thermal component as shown in Eq. \ref{eq: simple sed model}. The different coloured lines show the effect of increasing the \alphanth{} from the canonical value of $-0.8$ to up to $-1.5$. In both the panels the dashed horizontal lines show the value of \alphacs{}$=-0.64$ for $\fth=0.1$, and $\alphanth=-0.8$ from Eq. \ref{eq: simple sed model}. And the dashed vertical line shows the expected value of the thermal fraction at 3 and 6 GHz for such a radio spectrum. We also show our LzLCS observations on this plot in blue points. }
  \label{fig: specindex simplemodel}
\end{figure}

As seen in Fig. \ref{fig: total spec index fesc}, the observed spectral index across the C- and S-bands, \alphacs{}, flattens with increasing \fesc{}. This hints towards a flattening of \alphacs{} due to the young stellar ages of high \fesc{} galaxies (\fesc{} $> 0.1$). {Young stellar systems (ages $\leq 5$Myrs) would have a high thermal fraction due to a lack of significant SN activity (and thus nonthermal emission).} This will lead to a gradual flattening of \alphacs{} with decreasing stellar age. {In Fig. \ref{fig: specindex simplemodel}, we show the dependence of \alphacs{} on the estimated thermal fraction at 3 GHz (top panel) and 6 GHz (bottom panel) for our LzLCS sample.} We quantitatively show the effect of increasing the \fth{}, defined at a reference frequency of 3 GHz (top panel) and 6 GHz (bottom panel), on the observed \alphacs{}. This is calculated using the simple model for the radio-SED consisting of a thermal and a non-thermal component as described in Eq. \ref{eq: simple sed model}. The blue line shows the variation for a canonical value of $\alphanth=-0.8$. We also show the effect of steepening \alphanth{} to up to $-1.5$ in different coloured lines. The dashed horizontal line shows the value of \alphacs{}$=-0.64$ which is expected for a standard $\fth=0.1$ at 1 GHz, and $\alphanth=-0.8$. The dashed vertical line shows the value of the expected thermal fraction at 3 and 6 GHz for such a model.

As expected in all the models with both flat and steep values of \alphanth{}, \alphacs{} flattens as we increase \fth{}. Almost half of our sample lies within the region bracketed by these models. Thus their variation in the spectral indices in principle can be associated with their different values of \fth{} and \alphanth{}. However, very flat spectrum sources (\alphacs{} $> -0.5$) would need other mechanisms to further flatten the spectrum, for instance, FFA as discussed below. Similarly very steep spectrum sources (\alphacs{} $< -1.5$) would either need unusually steep \alphanth{}, or other mechanisms to steepen \alphanth{}, e.g., in a break model or CR escape which we discuss in Sect.  \ref{sec: non-leakers discussion}.


\begin{figure}
  \resizebox{\hsize}{!}{\includegraphics{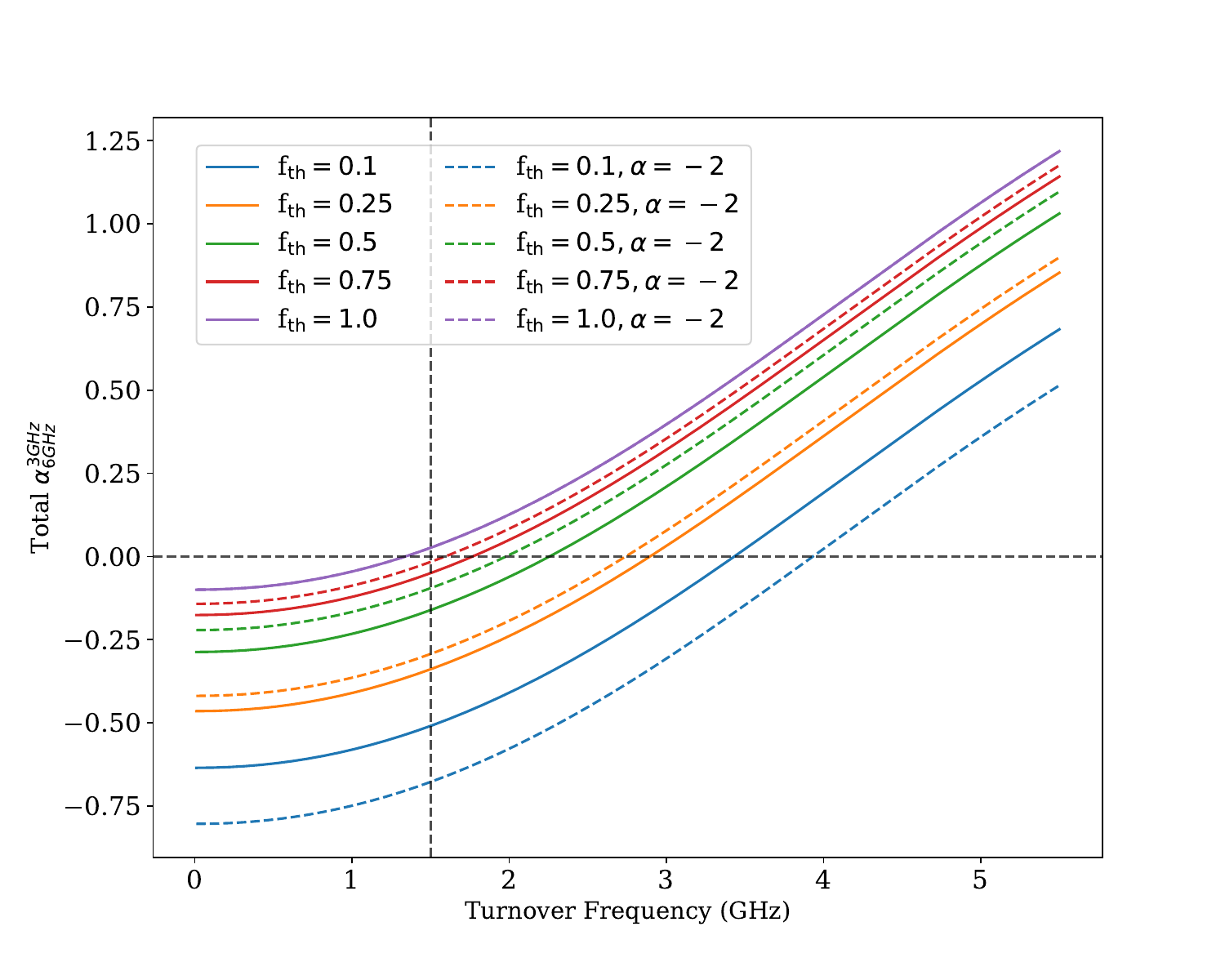}}
  \caption{Effect of increasing the turnover frequency (\nuturn{}) on \alphacs{}. The solid lines are for a canonical $\alphanth{}$=-0.8, with the blue line corresponding to a $\fth{} = 0.1$. The other lines shows the effect of increasing the \fth{}. The dashed lines are for the same range of \fth{} but with a more steep $\alphanth = -2$. The dashed horizontal line shows a value of \alphacs{}$= 0.$ This shows that we need FFA with a \nuturn{} of $\gtrsim 1.5$ GHz to flatten \alphacs{} $> 0$.
  }
  \label{fig: specindex turnover}
\end{figure}

Also, FFA has a strong effect on flattening the radio-SED. In normal galaxies, the effect of FFA is usually observed at low frequencies below $\sim 30$MHz \citep{Chyzy18}. For dense starbursts, FFA can be important at relatively higher frequencies around $\sim 700$MHz for starburst galaxies like in M82 \cite[see e.g.,][]{Chyzy18} and close to GHz for ULIRGS \citep[e.g.,][]{Clemens10, Galvin18}. However, for low metallicity BCDs, for example in SBS 0335-052, which host a dense super star cluster, \citet{Hunt04} observed a turnover frequency between 2-3 GHz, corresponding to an EM of $\sim 8 \times 10^7$pc cm$^{-6}$. In Fig. \ref{fig: specindex turnover}, we show the effect of increasing \nuturn{}, from 10 MHz to $\sim$5 GHz, on the observed \alphacs{}. We also show the effect of increasing the \fth{} from 0.1 to 1.0 on \alphacs{} shown in solid lines. In dashed lines, we show the same set of curves with the effect of steepening the \alphanth{} to $-2.0$. We notice that in order to flatten \alphacs{} $> 0.0$, in all the models we need a \nuturn{} $\gtrsim 1.5$GHz, which corresponds to an EM $\sim 1.2 \times 10^7$ pc cm$^{-6}$ assuming a $z \sim 0.3$ to convert the observed \nuturn{} to the rest-frame and the T$_\mathrm{e}$ of $10^{4}$ K. 
This is also in line with the correlation seen between \fesc{} and \sfrdensity{} \citep[e.g.,][]{Izotov16a, Naidu20, Flury22b}. Hence, increasing \sfrdensity{} (and thus the EM) can indirectly lead to an increase in \nuturn{} and thus flatten \alphacs{}. This can become particularly important for the high \fesc{} which has the highest \sfrdensity{}. {Finally, for normal star-forming galaxies in the KINGFISHER sample, it is observed that  \alphanth{} flattens with increasing \sfrdensity{} \citep{tabatabaei2017, Tabatabaei22}.} Such an effect is also observed in the arms and inner regions within a galaxy where the gas density is higher and the \alphanth{} can flatten to even higher than -0.5 \citep{Basu15}. More importantly in the case of IZw 18, a nearby low-metallicity BCD, \citet{Hunt05} finds a rather flat $\alphanth \approx -0.5$. {This is associated with the CR energy index being close to the injection spectrum during SN acceleration \citep{tabatabaei2017, Basu15} which is expected in galaxies with high SFRs. Alternatively, the high SFR can lead to an amplification of turbulent magnetic fields \citep[e.g.,][]{Gressel08, Gent23} which can scatter high-energy CRs before they lose their energy to synchrotron losses \citep{tabatabaei2017}. Thus leading to \revtext{a} flat non-thermal spectral index.} Our LzLCS sources are several orders of magnitude denser than KINGFISHER sample leading to the possibility that they too have relatively flat \alphanth{} which can overall flatten \alphacs{}.

We conclude that since high \fesc{} galaxies have a relatively young age of the starburst and relatively high \sfrdensity{} they can have a high thermal fraction,  EM, and possibly a flatter \alphanth{}. These effects can altogether flatten the \alphacs{} which can explain the observed trend for strong leakers in Fig. \ref{fig: total spec index fesc}. Finally, a flat spectrum source is not always a strong leaker source. We have observed a few outliers in the \alphacs{}-\fesc{} relation. {This indicates flat spectrum is possibly necessary but not a sufficient condition for identifying strong leakers. }




\subsection{Considering why non-leakers have a steep radio spectrum}
\label{sec: non-leakers discussion}

\begin{figure}
\centering
  \resizebox{\hsize}{!}{\includegraphics{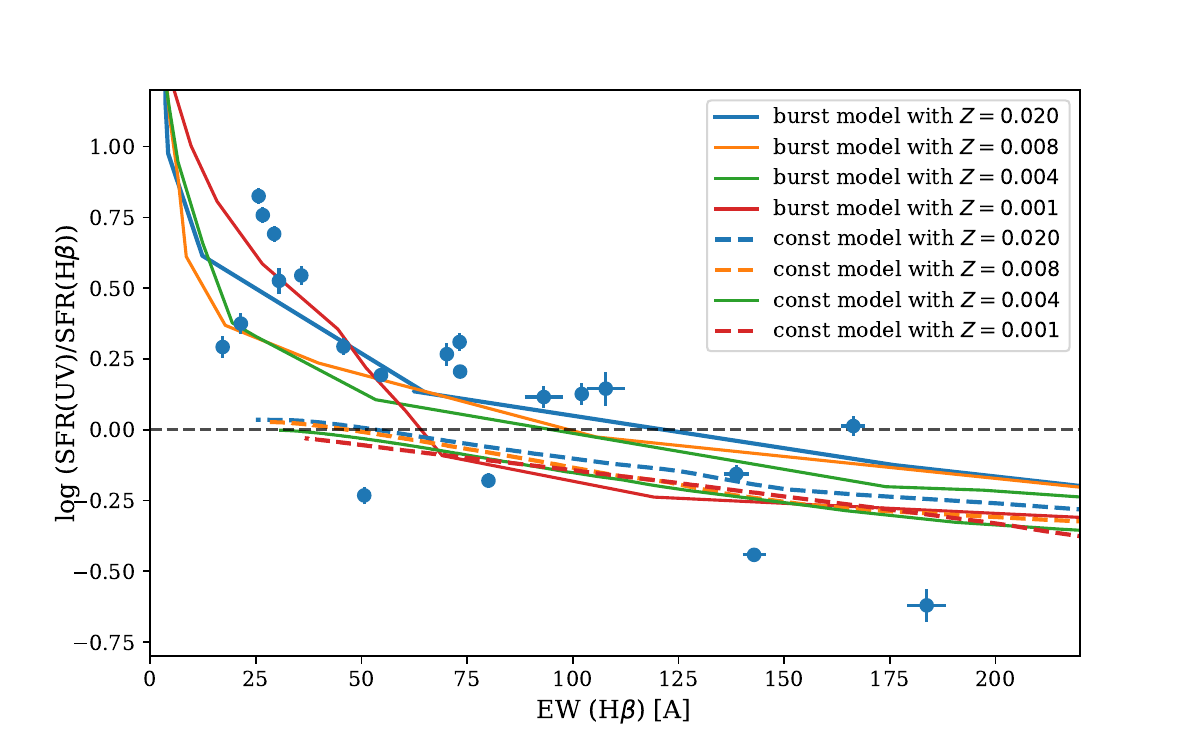}}
  \caption{Ratio between the SFR derived using UV and H$\beta$ luminosity as a function of the EW(H$\beta$). We note that typically galaxies with lower EW show a higher UV-based SFR compared to H$\beta$ based SFR. We also overplot the stellar population synthesis models from \cite{schaerer03} for a single burst model in solid lines, and continuous star-formation in dashed lines. The different colours blue, orange, green, and red represent the different metallicities corresponding to $Z=0.02$ (solar), 0.008, 0.004, and 0.001, respectively. }

  \label{fig: SFR UV Hbeta EW}
\end{figure}

\begin{figure}
  \resizebox{\hsize}{!}{\includegraphics{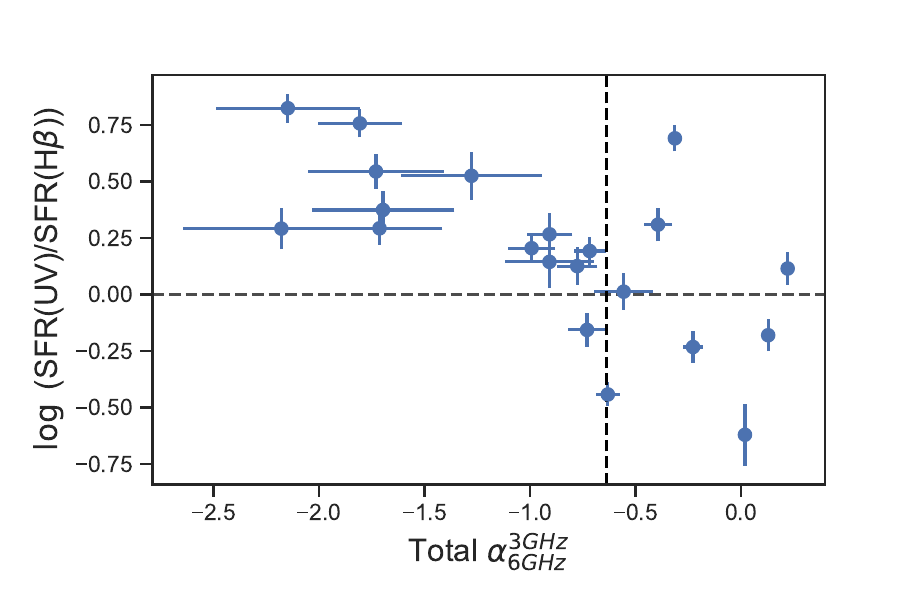}}
  \caption{Ratio between the SFR derived using UV and H$\beta$ luminosity vs. the \alphacs{} for our LzLCS sample. The horizontal line represents a ratio of 1. And the vertical line shows the \alphacs{} for a canonical radio spectrum. {We notice that galaxies with steep spectra have a higher UV-based SFR compared to H$\beta$ based SFR (and also lower \ewhbeta{} as seen in Fig. \ref{fig: SFR UV Hbeta EW})}.
  }
  \label{fig: SFR UV Hbeta specindex}
\end{figure}

\begin{figure}
  \resizebox{\hsize}{!}{\includegraphics{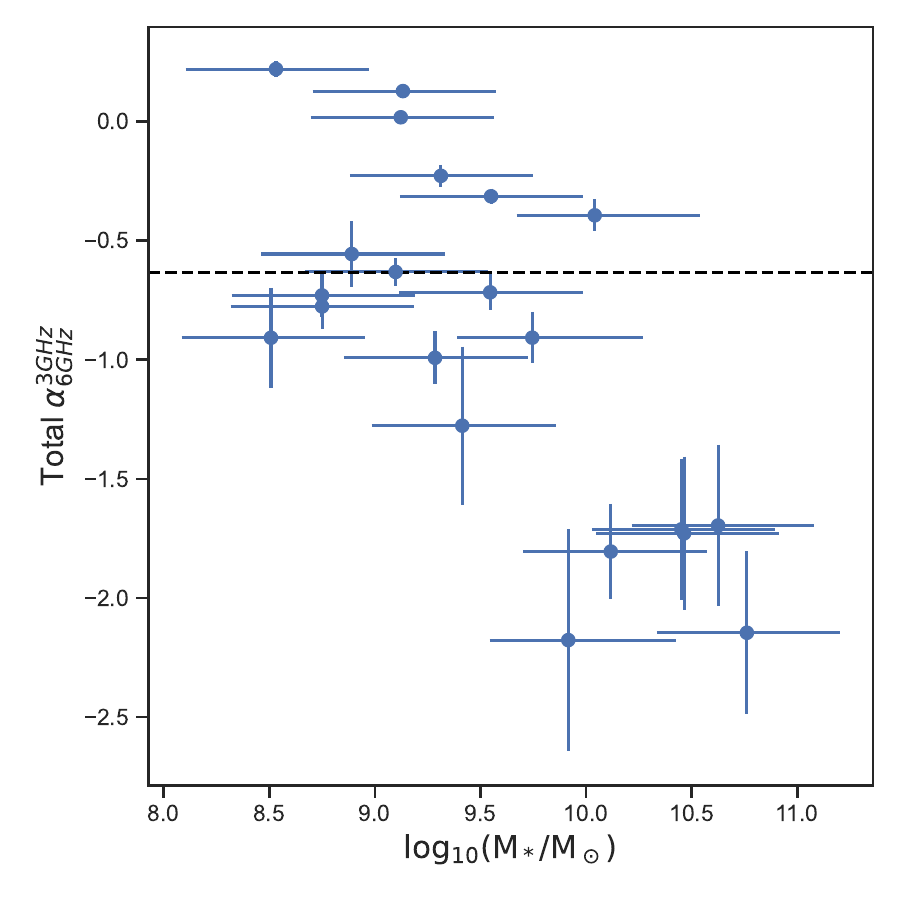}}
  \caption{{Dependance of \alphacs{} on the stellar mass of LzLCS sources. We notice that \alphacs{} steepens with increasing stellar mass. {The dashed line shows the \alphacs{} for a canonical radio spectrum. }}
  }
  \label{fig: specindex stellar mass}
\end{figure}

It is quite remarkable that the correlation between \alphacs{} and \fesc{} extends all the way towards non-leakers, and in particular, non-leakers have very steep \alphacs{} ($< -1$). Several processes related to either the complete loss of CRs or energy-dependent losses are usually invoked to explain the steepening of \alphacs{}, for example, a break, cutoff, or curvature at a higher frequency in the radio spectrum \citep[e.g., see the models listed in][]{klein2018}. A break in the spectrum is produced in an open-box model where the CRs can escape the halo of the galaxy \citep{lisenfeld2004}. This break occurs due to a high convective speed due to which the CRs can leave the halo \citep[e.g.,][]{lisenfeld2004, klein2018}. The break steepens \alphanth{} by 0.5 above a characteristic frequency termed as the break frequency \citep[][]{klein2018}. The break frequency is related to the convection speed and size of the halo. Thus, an original $\alphanth \sim -0.8$ can easily steepen to $\sim -1.3$ above the break frequency. A cutoff in a spectrum occurs when there is an abrupt change in the CR energy spectrum arising due to a relatively recent starburst a few Myrs ago. This happens in a so-called ``single-injection scenario,"  where most of the CRs were produced in a recent starburst. The high-energy CRs will lose their energy faster than the low-energy ones, and it will produce a strong break above some CR energy scale \citep{Schlickeiser84, klein2018}. This CR break energy is related to the break frequency and depends inversely on the time of CR injection \citep[see Eq. 4 in ][]{lisenfeld2004}.

Breaks in the radio spectrum were observed in NGC 1569, a nearby dwarf starburst, at around 8 GHz \citep{Israel88}, and several other nearby dwarf galaxies at GHz frequencies ranging from 1-12 GHz \citep{Deeg93, klein2018}. \citet{Israel88} applied the single-injection CR acceleration model to NGC 1569 and estimated the injection time to $\sim$5 Myrs ago. \cite{lisenfeld2004} has interpreted the same break in NGC 1569 due to a single-injection CR model (also termed as the cutoff model) or due to CR outflows, however, they favor the latter scenario. \citet{Deeg93} interpret several of their convex radio spectra in BCDs due to a break from a single-injection CR model. 


{ If non-leakers in the LzLCS sample are at the ``the low-phase of a bursty SF episode," namely, those} \  that recently stopped star formation or exhibit a declining SFR over time, we expect to see to first-order differences in their SFRs derived using UV and H$\beta$ luminosities, since the two trace SFR over long ($\sim$100 Myrs) and short timescales ($\sim$10 Myrs), respectively \citep[e.g.,][]{Kennicutt12}. In Fig. \ref{fig: SFR UV Hbeta EW} we compare the ratio \SFRUV/\SFRHbeta\ with the \ewhbeta{} for our sample. 
We notice that sources with the lowest \ewhbeta{} also have the highest ratio of \SFRUV{}/\SFRHbeta{}, which is naturally explained by relative age differences, with younger objects having higher \ewhbeta. To demonstrate this, we overplot predictions from the synthesis models of \cite{schaerer03} for metallicities bracketing those of the LzLCS and both for instantaneous burst models and constant SFR. Qualitatively the observed trend of increasing \SFRUV/\SFRHbeta\ with decreasing \ewhbeta, corresponding to a temporal evolution, is reproduced, and models of instantaneous burst (or other declining SF histories) also naturally lead to \SFRUV/\SFRHbeta$>1$.

In Fig. \ref{fig: SFR UV Hbeta specindex}, we compare this SFR ratio with the observed \alphacs{} for our LzLCS sample. We note  that the sources with the steepest spectra (thus non-leakers), show the highest ratio of \SFRUV/\SFRHbeta.  In such a cutoff model we can also estimate the CR injection timescale \citet[using Eq. 4 from][]{lisenfeld2004} for a break frequency close to 6 GHz and with magnetic fields ranging from 10-20 $\mu$G. We find that in such a model for the entire range of magnetic fields, a break can occur below 6 GHz within 10 Myrs. This timescale is consistent with the timescale on which we expect to see differences in the UV-based and H$\beta$-based SFR, and further supports our idea that the steepening of our LzLCS sources is because of their declining SFR with time. 

{Further, \citet{Heesen22} has shown a steepening of the radio spectral index with galaxy stellar mass for nearby galaxies in the LOw Frequency ARray (LOFAR) survey. {Their study which is at lower frequencies (144-1440 MHz) than ours, suggests that more massive galaxies have experienced a stronger synchrotron energy loss leading to a steeper radio spectra.} In Fig. \ref{fig: specindex stellar mass} we observe a similar trend for our LzLCS sources. In Table \ref{table: kendall tau} we also find a good correlation between stellar mass and \alphacs{}. Overall, it is difficult to distinguish which galaxy physical parameter drives the steepening of \alphacs{}. However, physically in both the scenarios, a significant loss of high-energy CRs can lead to the steepening of \alphacs{}.}

Alternatively, it is possible that the steepening is caused due to CR escape from advection or diffusion. In the intriguing case of the starburst dwarf galaxy, IC 10, there is evidence of SN feedback in the form of a non-thermal superbubble \citep{Yang93}. The spectral index is steep (close to -1.0) and is found to be consistent with CR escape due to advection or diffusion \citep{Hessen2011, Heesen18}. Strong outflows are also observed in compact starbursts \citep[e.g.,][]{Weiner09, Heckman11, Chisholm17}. Given the typically small sizes of these galaxies, the advection/diffusion timescales are expected to less than 1 Myr \citep{Sebastian19}. However, it is not directly very clear why the advection/diffusion driven CR escape should be more prominent in non-leakers as opposed to leakers. \citet{Farcy22} showed that the presence of CRs in simulated galaxies tend to smooth out the ISM thus filling it with denser gas and hence reducing \fesc{}. In the future, the inclusion of high and low-frequency radio data should better constrain the radio-SED and help us in distinguishing between the break or cutoff model for the steepening of \alphacs{}.


\subsection{Overall picture relating LyC escape and RC variations}
In summary, since the radio spectral index is a proxy for the age of the starburst \citep{Cannon04, Hirashita06}, we may possibly observe a relation between \alphacs{} and \fesc{}. Using a timescale argument we propose the following evolutionary picture. After the onset of a starburst (< 5 Myr) typically in dense ISM conditions, the radio spectrum will be flat (\alphacs{} $> -0.65$) due to i) thermally dominated radio emission as most of the young stars haven't yet undergone supernova ii)  the less dominant non-thermal emission which would be arising when some of the first supernovae start to explode will accelerate fresh CRs that have a flat non-thermal spectrum (> -0.65) and iii) the presence of dense ISM conditions can lead to a flat radio spectrum due to the strong effect of FFA at GHz frequencies. These physical conditions which lead to a flat \alphacs{} are also when the \fesc{} is highest. At later times (5-10 Myrs) when the starburst is declining along with the \fesc{}, high-energy CRs also lose their energies (in $\sim$10 Myrs) which results in the steepening of \alphacs{} ($< -0.65$). Consequently, the galaxies will quickly move down along the \alphacs{}-\fesc{} relation in Fig. \ref{fig: total spec index fesc} in around 10 Myrs. If there are any further bursts of star formation in these galaxies they can move along this sequence. Alternatively, the steepening can also be caused due to CR escape. For galaxies showing continuous star formation over 100 Myrs timescale, the CRs would have diffused in the galaxy and have a non-thermal spectral index close to the canonical value of -0.8 \citep{condon1992}. Such galaxies would then have a total \alphacs{} close to $-0.65$ and \fesc$<1$\%.  

Finally, the fact that some properties of the radio continuum correlate with the LyC escape fraction could indicate that the escape of ionizing photons is not highly directional, namely, which~ happens in a relatively isotropic manner. This is due to the fact that the observational determination of LyC escape (\fesc) measures a directional quantity, namely the escape of ionizing photons along the line-of-sight towards the observer, whereas RC emission is a priori a volumetric probe and orientation-independent. Therefore the finding of a correlation between \fesc\ and RC continuum properties would indicate that \fesc\ is also orientation-independent, to first order. This is consistent with previous results inferred from correlations between global galaxy properties and \fesc{} \citep{Wang19}.
Now in reality the relation between \alphacs{} and \fesc, for instance, shows a considerable scatter, which might partly be driven by some orientation-dependence. Better statistics, and a more fundamental understanding of the origin of these correlations will therefore be needed to properly address the question of non-isotropic LyC escape.
Other volumetric measures at other wavelengths also exist. For example, the nebular emission lines in the optical are thought to be essentially optically thin, and \fesc\ is found to correlate with \Oratio, which could again indicate relatively isotropic LyC emission. However, the scatter in these relations is also large \citep[see][]{Izotov21b,Flury22b}, and there is no unique and direct causal connection between LyC escape and \Oratio, which a priori depends on several parameters, such as the ionization parameter, metallicity, age of stellar population(s), etc., and \fesc\ \citep{Nakajima2013Ionization-Stat,Ramambason2020Reconciling-esc}.
Possible orientation and geometrical effects of LyC escape are so far difficult to establish observationally from the available low-$z$ galaxy samples.

\section{Conclusions and future work}
\label{sec: conclusions}


 For the first time, we studied the radio continuum (RC) properties of galaxies with LyC measurements  based on VLA C- and S-band observations for a sample of 53 galaxies. We also obtained L-band observations for 17 of these sources. These galaxies were selected from the recent  Low-$z$ Lyman Continuum Survey \citep[LzLCS;][]{Flury22a,Flury22b}, which contains a large sample of low-$z$ ($z \sim 0.3-0.4$) star-forming galaxies with LyC measurements. We detected the RC at both C- and S-bands for about half of our sample, where we were then able to measure the total spectral index (\alphacs{}) and thermal fraction. 

{Our study highlights an interesting connection between the radio spectral index and the escape of LyC photons. Overall, we associate this with the rapidly changing SFH (on a few Myr timescale) due to the onset of starburst and its decline in LzLCS galaxies. These changes in SFHs are traced by the CR energy spectrum (and, thus, by the radio spectral index) and they also have an imprint on the LyC escape fraction.} 
Our main results can be summarized as follows.

   \begin{enumerate}
      \item The RC spectral index between C- and S-bands (\alphacs{}) for LzLCS sources shows a \revtext{large variation} from that of typical star-forming galaxies, with sources showing a very flat (> -0.1) to very steep spectral index (< -1).
      \item We find that LzLCS galaxies have a normal to significant contribution of the non-thermal component to the total radio emission.
      This provides the first direct evidence of SN activity in these galaxies and suggests that SN can contribute to the escape of LyC photons.
      \item We found a positive correlation between \alphacs{} and the escape fraction of Lyman continuum photons, \fesc; thus, strong leakers have flat \alphacs{} and non-leakers exhibit a very steep \alphacs{}. 
      \item We argue that a combination of young ages (high \fth{}), FFA with a turnover at GHz frequencies and CRs with a flat spectrum (young) can lead to the flattening of \alphacs{}. Such physical conditions typical of young and dense starbursts are also seen when galaxies are expected to exhibit the highest \fesc{}. 
      \item  Steep spectrum sources are associated with low \fesc{}. They show evidence of declining SFRs on short timescales and with higher stellar masses, both of which may lead to strong synchrotron losses and   a steeper radio spectrum as a result. 
      \item Our work highlights the evolutionary nature of \fesc{}, on short timescales ($\sim 10$ Myrs,) and its association with SNs, the CR energy spectrum, and magnetic fields related to the escape of LyC photons.
      \item {The RC-SFR relation of LzLCS sources show a large scatter. Using RC as an SFR tracer for such galaxies is challenging when only a single or narrow frequency range is available.} 
      \item {The \SFRRC{} shows a systematic offset from the \SFRRC{6 GHz}-\SFRUV~(\SFRHbeta)}{ relation for local normal star-forming galaxies from \citet{tabatabaei2017}. We calibrated the \SFRUV{} (and \SFRHbeta{})} {for a combination of \SFRRC{} and \alphacs{} values, which can be used to estimate their SFRs.} Information on the radio-SED will be crucial to {robustly} calibrate the radio-based SFRs in such galaxies. Since these galaxies are analogs of high-$z$ galaxies ($z > 3$), our study gives important insights into such galaxies. This is particularly relevant in the context of the upcoming highly sensitive radio facilities, such as SKA and ngVLA, aimed at detecting star-forming galaxies at high redshifts. 
   \end{enumerate}

{In our current study, using just the C- and S-band data, we cannot robustly estimate all of the physical parameters we are looking to obtain. In the future, we plan to conduct low-frequency (< 1.5 GHz) and high-frequency (> 8 GHz) follow-up observations of LzLCS sources to better study their radio-SEDs. This will allow us to fit complex radio-SED models, which will be useful in accurately estimating the non-thermal spectral index and the thermal fraction. This will also help us estimate some of the physical parameters that are lacking; for example, the SN rate, CR energy spectrum, and magnetic fields, and their relation to LyC escape. These approaches will improve our understanding of the  role of SN feedback in the escape of LyC photons. Finally, our current radio follow-up of LzLCS has missed a number of strong leakers. This will be the subject of a follow-up study in the future. }

\begin{acknowledgements}
We thank the referee for their comments which helped improve the draft. O. Bait (OB) is supported by the {\em AstroSignals} Sinergia Project funded by the Swiss National Science Foundation. OB would like to thank Jennifer Schober and Aayush Saxena for useful comments on the manuscript. OB would also like to thank Drew Medlin and the NRAO science helpdesk team for their extensive help with running the VLA pipeline and with the VLA data reduction. We would like to thank Brad Koplitz for his help with setting up some of the VLA scheduling blocks. ASL acknowledges support from Knut and Alice Wallenberg Foundation

Funding for the Sloan Digital Sky 
Survey IV has been provided by the 
Alfred P. Sloan Foundation, the U.S. 
Department of Energy Office of 
Science, and the Participating 
Institutions. 

SDSS-IV acknowledges support and 
resources from the Center for High 
Performance Computing  at the 
University of Utah. The SDSS 
website is www.sdss4.org.

SDSS-IV is managed by the 
Astrophysical Research Consortium 
for the Participating Institutions 
of the SDSS Collaboration including 
the Brazilian Participation Group, 
the Carnegie Institution for Science, 
Carnegie Mellon University, Center for 
Astrophysics | Harvard \& 
Smithsonian, the Chilean Participation 
Group, the French Participation Group, 
Instituto de Astrof\'isica de 
Canarias, The Johns Hopkins 
University, Kavli Institute for the 
Physics and Mathematics of the 
Universe (IPMU) / University of 
Tokyo, the Korean Participation Group, 
Lawrence Berkeley National Laboratory, 
Leibniz Institut f\"ur Astrophysik 
Potsdam (AIP),  Max-Planck-Institut 
f\"ur Astronomie (MPIA Heidelberg), 
Max-Planck-Institut f\"ur 
Astrophysik (MPA Garching), 
Max-Planck-Institut f\"ur 
Extraterrestrische Physik (MPE), 
National Astronomical Observatories of 
China, New Mexico State University, 
New York University, University of 
Notre Dame, Observat\'ario 
Nacional / MCTI, The Ohio State 
University, Pennsylvania State 
University, Shanghai 
Astronomical Observatory, United 
Kingdom Participation Group, 
Universidad Nacional Aut\'onoma 
de M\'exico, University of Arizona, 
University of Colorado Boulder, 
University of Oxford, University of 
Portsmouth, University of Utah, 
University of Virginia, University 
of Washington, University of 
Wisconsin, Vanderbilt University, 
and Yale University.
\end{acknowledgements}

%
%

\bibliographystyle{aa} 
\bibliography{references.bib}

\begin{appendix} 

\section{Observation details of VLA S-, C-, and L-band sources in our sample.} \label{appendix sec: VLA obs details}
In Table \ref{table: obs detail}, we describe the details of the C- and S-band observations for our sources. And Table \ref{table: obs detail L band} shows the details of the L-band observations for a sub-sample of these sources.

\begin{table*}
\caption{VLA C- and S-band observation details.} 
\begin{tabular}{cccccc} 
\hline\hline \\
\label{table: obs detail}
Object & Obs. date & Flux cal & Phase cal & Phase center Ra & Phase center Dec \\ \\ \hline  \\     
J004743+015440 & 2021-11-17 & 3C 147 & J0059+0006 & 00h47m42.8s & +01d54m39.9s \\
J011309+000223 & 2021-12-08 & 3C 147 & J0059+0006 & 01h13m08.9s & +00d02m23.4s \\
J012910+145935 & 2021-10-07 & 3C 147 & J0204+1514 & 01h29m10.2s & +14d59m34.7s \\
J072326+414608 & 2021-11-05 & 3C 147 & J0713+4349 & 07h23m25.6s & +41d46m07.8s \\
J081112+414146 & 2022-02-03 & 3C 147 & J0818+4222 & 08h11m12s & +41d41m45.9s \\
J081409+211459 & 2021-12-01 & 3C 147 & J0842+1835 & 08h14m08.8s & +21d14m59.2s \\
J082652+182052 & 2022-01-02 & 3C 147 & J0842+1835 & 08h26m51.8s & +18d20m51.8s \\
J083440+480541 & 2022-02-15 & 3C 147 & J0818+4222 & 08h34m40.1s & +48d05m40.9s \\
J090918+392925 & 2022-02-15 & 3C 147 & J0818+4222 & 09h09m18.4s & +39d29m24.7s \\
J091113+183108 & 2021-12-30 & 3C 147 & J0854+2006 & 09h11m13.3s & +18d31m08.2s \\
J091207+523960 & 2022-02-08 & 3C 147 & J0921+6215 & 09h12m07.2s & +52d39m59.9s \\
J091208+505009 & 2021-11-25 & 3C 147 & J0834+5534 (S), J0832+4913 (C) & 09h12m08.1s & +50d50m08.6s \\
J091703+315221 & 2022-01-21 & 3C 147 & J0956+2515 & 09h17m02.5s & +31d52m20.6s \\
J092552+395714 & 2022-02-14 & 3C 147 & J0920+4441 & 09h25m51.9s & +39d57m13.7s \\
J094001+593244 & 2022-01-27 & 3C 147 & J0921+6215 & 09h40m00.7s & +59d32m44.4s \\
J095236+405249 & 2022-01-12 & 3C 147 & J0920+4441 & 09h52m35.9s & +40d52m48.8s \\
J095700+235709 & 2021-11-06 & 3C 147 & J0956+2515 & 09h57m00.4s & +23d57m09s \\
J095838+202508 & 2022-01-06 & 3C 147 & J0956+2515 & 09h58m38.4s & +20d25m07.6s \\
J101401+523251 & 2022-01-22 & 3C 286 & J1035+5628 & 10h14m01.5s & +52d32m50.7s \\
J102615+633308 & 2022-02-28 & 3C 286 & J1035+5628 & 10h26m15.2s & +63d33m08.5s \\
J103344+635317 & 2022-03-01 & 3C 286 & J1035+5628 & 10h33m44.1s & +63d53m17.2s \\
J103816+452718 & 2022-02-19 & 3C 286 & J0948+4039 & 10h38m16.2s & +45d27m17.8s \\
J105117+474357 & 2022-02-19 & 3C 286 & J1035+5628 & 10h51m16.9s & +47d43m57s \\
J105331+523753 & 2022-02-28 & 3C 286 & J1035+5628 & 10h53m30.8s & +52d37m52.9s \\
J110452+475204 & 2022-03-03 & 3C 286 & J1035+5628 & 11h04m52.2s & +47d52m04.3s \\
J112224+412052 & 2022-02-17 & 3C 286 & J1146+3958 & 11h22m24.3s & +41d20m52.4s \\
J112848+524509 & 2022-02-20 & 3C 286 & J1035+5628 & 11h28m47.6s & +52d45m09.4s \\
J112933+493525 & 2022-02-22 & 3C 286 & J1146+3958 & 11h29m32.8s & +49d35m24.6s \\
J113304+651341 & 2022-02-20 & 3C 286 & J1313+6735 & 11h33m03.8s & +65d13m41.4s \\
J115855+312559 & 2022-03-02 & 3C 286 & J1146+3958 & 11h58m54.8s & +31d25m59.2s \\
J115959+382422 & 2022-02-27 & 3C 286 & J1146+3958 & 11h59m58.9s & +38d24m22s \\
J120934+305326 & 2022-03-03 & 3C 286 & J1215+3448 & 12h09m34.5s & +30d53m26.2s \\
J121915+453930 & 2022-02-22 & 3C 286 & J1146+3958 & 12h19m14.7s & +45d39m29.6s \\
J123519+063556 & 2022-02-22 & 3C 286 & J1254+1141 & 12h35m18.5s & +06d35m56.4s \\
J124423+021540 & 2021-11-15 & 3C 286 & J1246-0730 & 12h44m23.4s & +02d15m40.5s \\
J124619+444902 & 2022-02-15 & 3C 286 & J1227+3635 & 12h46m19.5s & +44d49m02.4s \\
J124835+123403 & 2022-03-03 & 3C 286 & J1254+1141 & 12h48m34.6s & +12d34m02.9s \\
J124911+464535 & 2022-02-21 & 3C 286 & J1219+4829 & 12h49m11.2s & +46d45m34.5s \\
J125718+410221 & 2021-11-19 & 3C 286 & J1310+3220 & 12h57m18.3s & +41d02m21.4s \\
J130128+510451 & 2022-02-21 & 3C 286 & J1219+4829 & 13h01m28.3s & +51d04m51.2s \\
J131037+214817 & 2022-02-27 & 3C 286 & J1327+2210 & 13h10m36.7s & +21d48m17s \\
J131419+104739 & 2022-02-27 & 3C 286 & J1347+1217 & 13h14m18.9s & +10d47m39s \\
J131904+510309 & 2021-12-16 & 3C 286 & J1219+4829 & 13h19m03.8s & +51d03m09.3s \\
J132633+421824 & 2022-02-21 & 3C 286 & J1310+3220 & 13h26m33.4s & +42d18m24.1s \\
J132937+573315 & 2022-02-28 & 3C 286 & J1400+6210 & 13h29m37.4s & +57d33m15.2s \\
J134559+112848 & 2022-03-02 & 3C 286 & J1347+1217 & 13h45m58.9s & +11d28m47.8s \\
J141013+434435 & 2021-12-27 & 3C 286 & J1416+3444 & 14h10m13s & +43d44m35.1s \\
J144010+461937 & 2022-02-21 & 3C 286 & J1349+5341 & 14h40m09.9s & +46d19m37s \\
J151707+370512 & 2022-02-22 & 3C 286 & J1602+3326 & 15h17m07.4s & +37d05m12.3s \\
J155945+403325 & 2022-02-20 & 3C 286 & J1613+3412 & 15h59m44.6s & +40d33m25.3s \\
J160437+081959 & 2022-03-03 & 3C 286 & J1557-0001 & 16h04m36.7s & +08d19m59.1s \\
J164607+313054 & 2022-02-22 & 3C 286 & J1613+3412 & 16h46m06.5s & +31d30m53.5s \\
J164849+495751 & 2021-12-24 & 3C 286 & J1739+4737 & 16h48m49.4s & +49d57m50.9s \\
            \hline
\end{tabular}
    
\end{table*}

\begin{table*}
\caption{VLA L-band observation details.} 
\begin{tabular}{cccccc} 
\hline\hline \\
\label{table: obs detail L band}
Object & Obs. Date & Flux Cal & Phase Cal & Phase Center Ra & Phase Center Dec \\ \\ \hline  \\     
J004743+015440 & 2022-01-28 & 3C 147 & J0059+0006 & 00h47m42.8s & +01d54m39.9s \\
J011309+000223 & 2022-01-29 & 3C 147 & J0059+0006 & 01h13m08.9s & +00d02m23.4s \\
J012910+145935 & 2022-01-17 & 3C 147 & J0204+1514 & 01h29m10.2s & +14d59m34.7s \\
J072326+414608 & 2022-01-28 & 3C 147 & J0713+4349 & 07h23m25.6s & +41d46m07.8s \\
J081112+414146 & 2022-02-21 & 3C 147 & J0818+4222 & 08h11m12s & +41d41m45.9s \\
J081409+211459 & 2022-01-28 & 3C 147 & J0842+1835 & 08h14m08.8s & +21d14m59.2s \\
J082652+182052 & 2022-02-21 & 3C 147 & J0842+1835 & 08h26m51.8s & +18d20m51.8s \\
J083440+480541 & 2022-01-19 & 3C 147 & J0818+4222 & 08h34m40.1s & +48d05m40.9s \\
J090918+392925 & 2022-02-22 & 3C 147 & J0818+4222 & 09h09m18.4s & +39d29m24.7s \\
J091113+183108 & 2022-01-30 & 3C 147 & J0854+2006 & 09h11m13.3s & +18d31m08.2s \\
J091207+523960 & 2022-01-21 & 3C 147 & J0921+6215 & 09h12m07.2s & +52d39m59.9s \\
J091208+505009 & 2022-01-24 & 3C 147 & J0834+5534 & 09h12m08.1s & +50d50m08.6s \\
J091703+315221 & 2022-01-30 & 3C 147 & J0956+2515 & 09h17m02.5s & +31d52m20.6s \\
J092552+395714 & 2022-01-25 & 3C 147 & J0920+4441 & 09h25m51.9s & +39d57m13.7s \\
J095236+405249 & 2022-02-18 & 3C 147 & J0920+4441 & 09h52m35.9s & +40d52m48.8s \\
J095700+235709 & 2022-02-27 & 3C 147 & J0956+2515 & 09h57m00.4s & +23d57m09s \\
J095838+202508 & 2022-02-28 & 3C 147 & J0956+2515 & 09h58m38.4s & +20d25m07.6s \\
            \hline
\end{tabular}
    
\end{table*}

\section{VLA S- and C-band images for the detections in our sample.} \label{appendix sec: VLA maps}
\begin{figure*}
  \resizebox{\hsize}{!}{\includegraphics{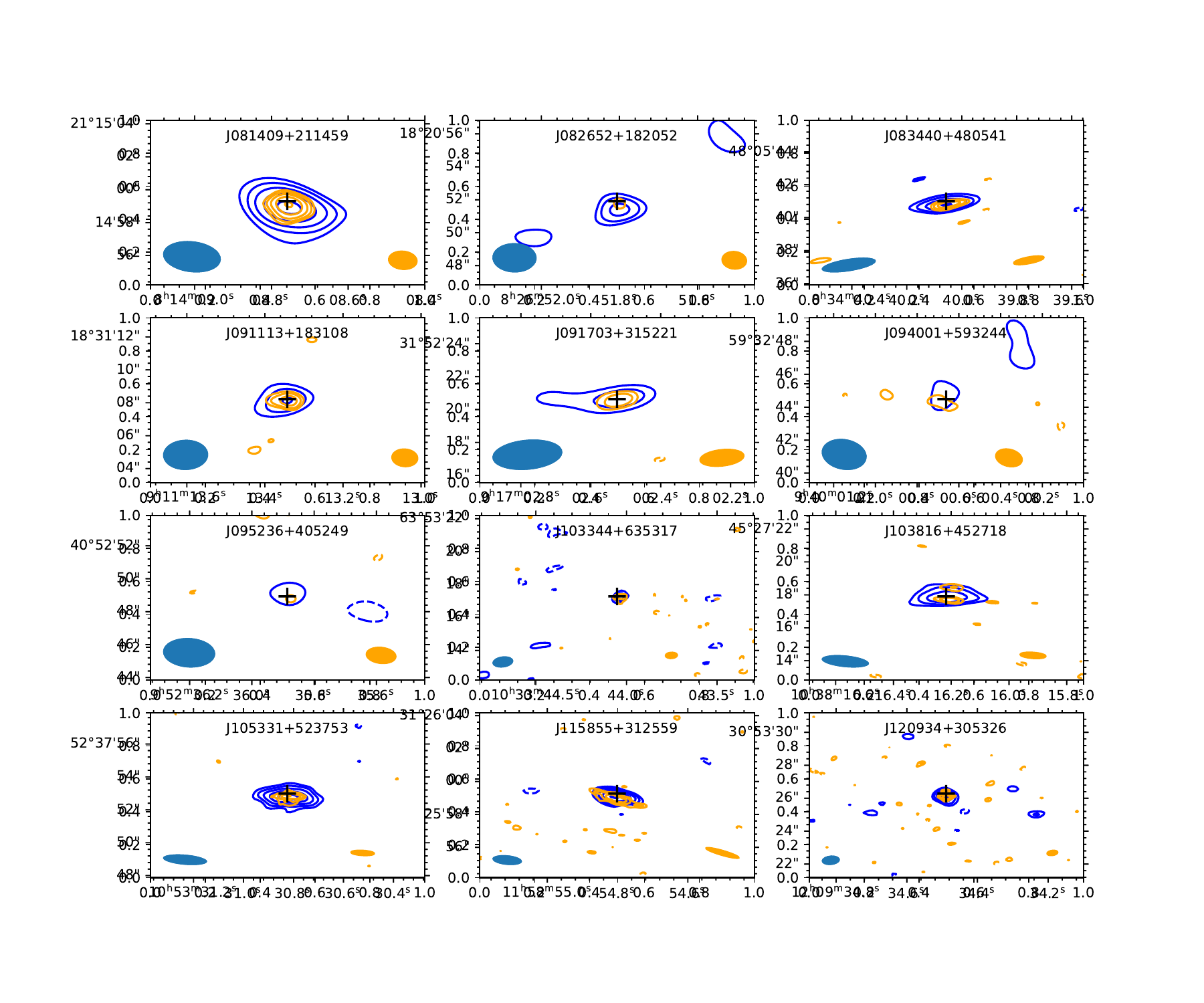}}
  \caption{ RC images for all the LzLCS detections in S-band (blue contours) and C-band (orange contours). The black crosshair represents the optical centre. The synthesized beam  is shown blue (orange) filled ellipticals for S-band (C-band). The contours start from 3-$\sigma$, and successively increase in powers of $\sqrt{2}$. The dashed contours represent the -3-$\sigma$ contour.}
  \label{fig: VLA maps 1}
\end{figure*}

\begin{figure*}
  \resizebox{\hsize}{!}{\includegraphics{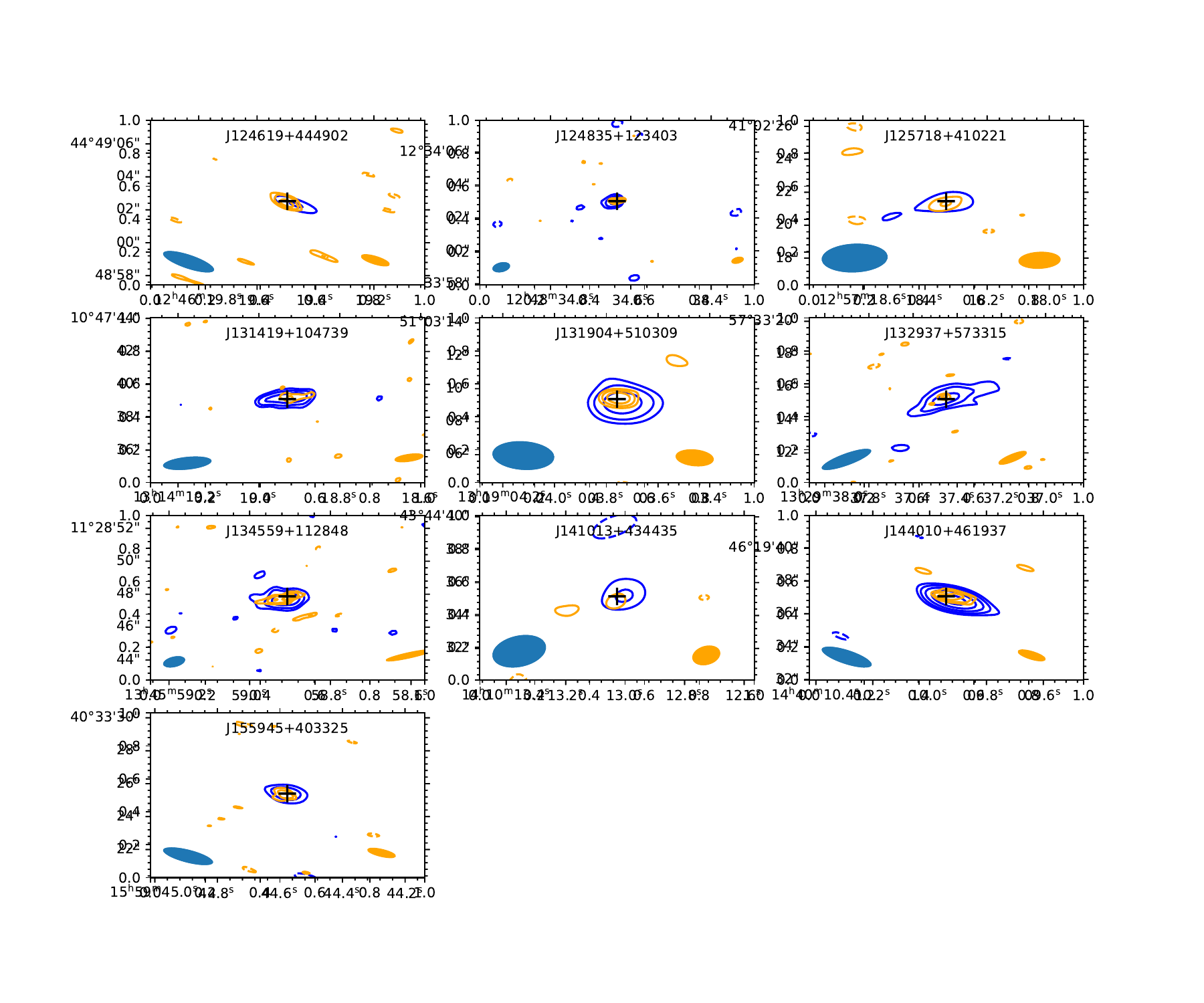}}
  \caption{ Continued from Fig. \ref{fig: VLA maps 1}.}
  \label{fig: VLA maps 2}
\end{figure*}
 In Figs. \ref{fig: VLA maps 1} and \ref{fig: VLA maps 2}, we show the VLA C-band (in orange) and S-band (in blue) contours overlaid on each other for our detection sample. 

\section{Radio-SED for our LzLCS sample at GHz frequencies} \label{appendix sec: All SEDs}
In Figs. \ref{fig: radio-SED 1} and \ref{fig: radio-SED 2}, we present the radio-SED across C-, S-, and also L-band (for those with C, S-band detections, and L-band observations.).
\begin{figure*}
  \resizebox{\hsize}{!}{\includegraphics{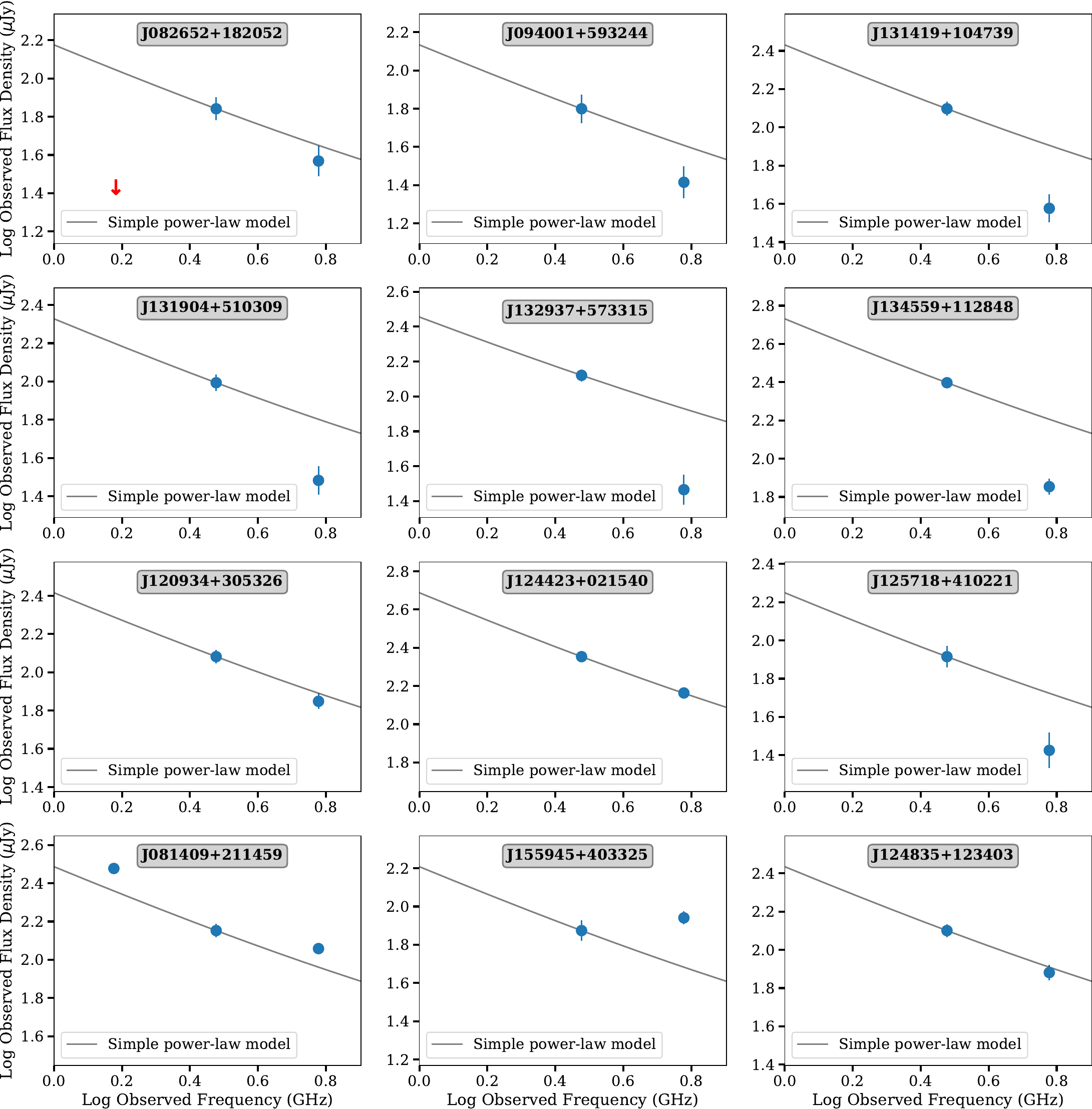}}
  \caption{ Same as Fig. \ref{fig: radio sed} for the remaining LzLCS sample with C- and S-bands detections. The red arrows represent the 3-$\sigma$ upper limits for L-band non-detections in our sample.}
  \label{fig: radio-SED 1}
\end{figure*}

\begin{figure*}
  \resizebox{\hsize}{!}{\includegraphics{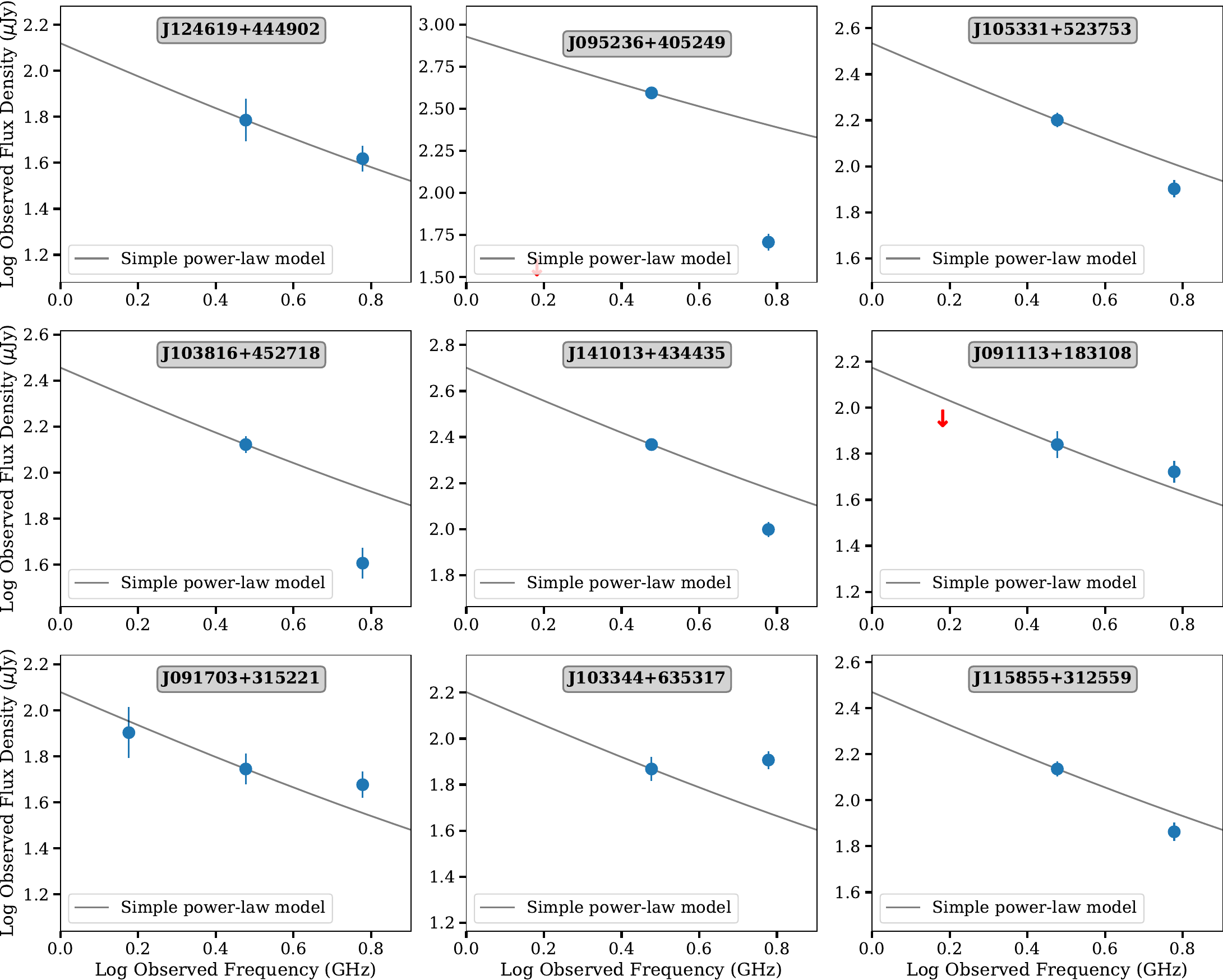}}
  \caption{Continued from Fig. \ref{fig: radio-SED 1}.}
  \label{fig: radio-SED 2}
\end{figure*}

\section{{Possible contamination of the radio flux of J130128+510451 from a neighbouring AGN jet}}
\label{appendix sec: J130128_AGN_jet}
\begin{figure*}
  \resizebox{\hsize}{!}{\includegraphics{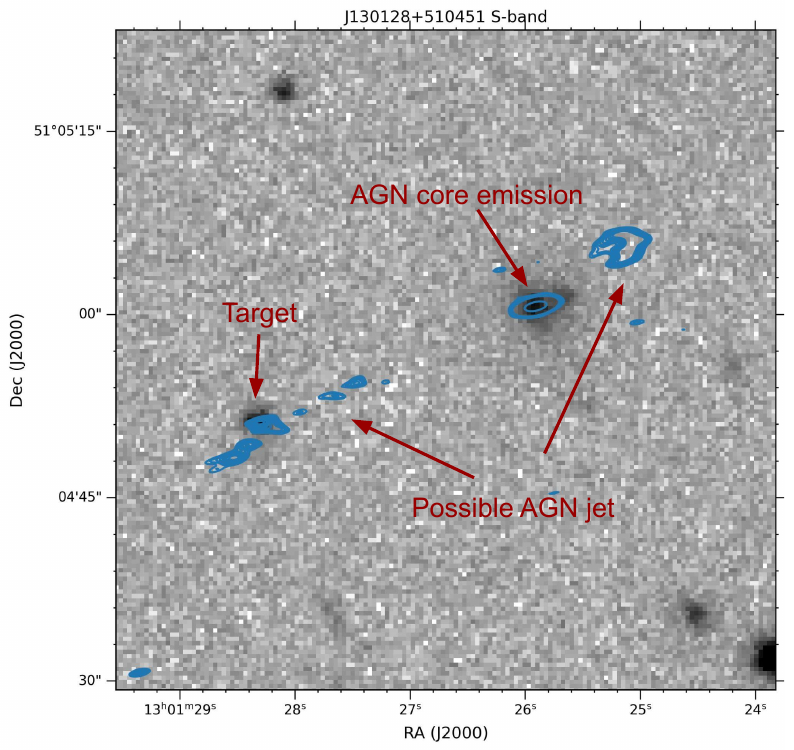}}
  \caption{{Large-scale VLA S-band image around the target, J130128+510451. The radio emission from the target seems to be contaminated by the jet emission from a nearby radio-loud AGN. The radio contour levels start from 3.5$\sigma$ and rise in powers of 1.2. We also added three more contours at 0.1, 1.0, and 2.0 mJy/beam to show the bright core emission from the neighbouring source, J130126+510500.}}
  \label{fig: J130128_AGN_jet}
\end{figure*}

{In Fig. \ref{fig: J130128_AGN_jet}, we show the large-scale VLA S-band image around J130128+510451. We notice that a neighbouring radio-loud source (J130126+510500: RA: 13h01m25.908s; DEC: $+$51d05m00.66s) shows a strong jet-like emission along the north-west direction. It possibly has diffuse extensions along the south-east direction. This latter extension coincidentally passes directly through our target, thus potentially contaminating the radio flux of our target. However, deeper radio data is required to confirm this diffuse south-east jet-like feature and its association to J130126+510500.} 

\section{{Details on the \SFRRC{}, \alphacs{} fit to SFR from other tracers}}
\label{appendix sec: SFR fit details}

{ We perform a linear fit between $\log$ \SFRRC{}, \alphacs{}, and the $\log$ SFR measured using the UV flux and H$\beta$ line flux. Here, the fit is performed separately for \SFRRC{3 GHz} and \SFRRC{6 GHz}, to \SFRUV{} and \SFRHbeta{}.}  {Since all three variables have error bars on them, we use the bootstrap method to take into account the uncertainties. We draw 100 random samples for each of the three variables assuming a Gaussian distribution with the 1-$\sigma$ errors on each variable representing the width of the Gaussian. This leads to a million combinations of variable pairs (a pair consists of three variables, two SFRs, and \alphacs{}). We performed a linear fit on each of these pairs using Eq. \ref{eq: SFR RC fit}. The best-fit parameters and the uncertainty are estimated by taking the mean and standard deviation of each of the fit parameters from the entire sample. The residual in the fit is the mean of the residual from the entire sample. Table \ref{table: SFR RC fit parameters} provides the fit parameters, corresponding errors, and the residual from the fit.}
\end{appendix}
\end{document}